\documentclass[11pt]{article}
\usepackage{jcapmod}
\usepackage{rotating}
\usepackage{amsmath}
\usepackage{mathtools}
\usepackage{braket}
\usepackage{amsthm}
\usepackage{amsfonts}
\usepackage{graphicx}
\usepackage{psfrag}
\usepackage{hyperref}
\usepackage{amssymb}
\usepackage{lscape}
\usepackage{slashed}
\usepackage{lscape}
\usepackage{color}
\usepackage{url}
\usepackage{caption}
\usepackage{bm}
\usepackage{subfig}
\usepackage{epsfig}
\usepackage{soul}
\usepackage{physics}

\def\trh{T_{\rm RH}}

\newcommand{\beq}{\begin{equation}}
\newcommand{\eeq}{\end{equation}}
\newcommand{\bea}{\begin{eqnarray}}
\newcommand{\eea}{\end{eqnarray}}
\newcommand{\ga}{\lower.7ex\hbox{$\;\stackrel{\textstyle>}{\sim}\;$}}
\newcommand{\la}{\lower.7ex\hbox{$\;\stackrel{\textstyle<}{\sim}\;$}}

\newcommand*\diff{\mathop{}\!\mathrm{d}}

\newcommand{\arh}{a_\text{RH}}

\newcommand{\aend}{a_\text{end}}

\def\del{\partial}
\def\lrd{\overset{\leftrightarrow}{\del}}

\makeatletter
\newcommand{\Cen}[2]{%
  \ifmeasuring@
    #2%
  \else
    \makebox[\ifcase\expandafter #1\maxcolumn@widths\fi]{$\displaystyle#2$}%
  \fi
}
\makeatother

\begin{document}

\begin{flushright}
UMN--TH--4303/23, FTPI--MINN--23/22, OU--HET--1209
\end{flushright}

\vspace{0.5cm}
\begin{center}
{\bf {\Large The Role of Vectors in Reheating}}
\end{center}

\vspace{0.05in}
\begin{center}
{\bf Marcos~A.~G.~Garcia}$^{a}$,
{\bf Kunio Kaneta$^{b}$},
{\bf Wenqi Ke$^{c,d}$},
{\bf Yann Mambrini}$^{e}$,
{\bf Keith~A.~Olive}$^{c}$, 
{\bf Sarunas Verner$^{f}$}
\end{center}

\begin{center}
 {\em $^a$Departamento de F\'isica Te\'orica, Instituto de F\'isica, Universidad Nacional Aut\'onoma de M\'exico, Ciudad de M\'exico C.P. 04510, Mexico}\\[0.2cm] 
 {\em ${}^b $ Department of Physics, Osaka University, Toyonaka, Osaka 560-0043, Japan}
\\ {\em ${}^c$ Sorbonne Universit\'e, CNRS, Laboratoire de Physique Th\'eorique et Hautes Energies, LPTHE, F-75005 Paris, France.
}\\
 {\em $^d$William I. Fine Theoretical Physics Institute, School of
 Physics and Astronomy, University of Minnesota, Minneapolis, MN 55455,
 USA}\\[0.2cm] 
  {\em $^e$ Universit\'e Paris-Saclay, CNRS/IN2P3, IJCLab, 91405 Orsay, France}\\[0.2cm] 
{\em ${}^f$Institute for Fundamental Theory, Physics Department, University of Florida, Gainesville, FL 32611, USA}

\end{center}

\bigskip

\centerline{\bf ABSTRACT}
We explore various aspects concerning the role of vector bosons during the reheating process. Generally, reheating occurs during the period of oscillations of the inflaton condensate and the evolution of the radiation bath depends on the inflaton equation of state. For oscillations about a quadratic minimum, the equation of state parameter, $w = p/\rho =0$, and the evolution of the temperature, $T(a)$ with respect to the scale factor is independent of the spin of the inflaton decay products. However, for cases when $w>0$, there is a dependence on the spin, and here we consider the evolution when the inflaton decays or scatters to vector bosons. We also investigate the gravitational production of vector bosons as potential dark matter candidates. Gravitational production predominantly occurs through the longitudinal mode. We compare these results to the gravitational production of scalars. 

\noindent

\vspace{0.2in}

\begin{flushleft}
November 2023
\end{flushleft}
\medskip
\noindent

\newpage

\hypersetup{pageanchor=true}

\tableofcontents

\newpage
\section{Introduction}
While the inflationary paradigm \cite{reviews} is well-embedded into standard Big Bang cosmology, it lies beyond the Standard Model (SM) framework. Consequently, there is considerable model dependence regarding how the inflationary sector couples to the Standard Model—couplings that are necessary for establishing an early period of radiation domination. It is often assumed that the inflaton must decay or scatter to reheat the universe \cite{dg,nos}. This transfer of energy typically occurs after the period of accelerated expansion has ended, and the inflaton begins to oscillate around its potential minimum. The thermalization of these decay or scattering products is responsible for generating the thermal background that eventually comes to dominate the energy density. The process of establishing this thermal background is not instantaneous \cite{Giudice:2000ex,Bernal:2020gzm,GMOP}, and the reheating temperature, along with the scaling of the radiation density during reheating, depends on the spin of the final state particle and the shape of the potential near the minimum that governs inflaton oscillations \cite{GKMO1,GKMO2,Becker:2023tvd}.

Indeed, when the inflaton potential around its minimum is dominated by a quadratic term,
$V(\phi) \propto \phi^2$, the energy density of the inflaton, $\rho_\phi$, scales as $\rho_\phi \propto a^{-3}$, where $a$ represents the cosmological scale factor. As oscillations begin, and assuming that the inflaton decays, a bath of relativistic particles is produced which subsequently thermalizes,\footnote{We do not address the problem of thermalization in this paper. For discussions on this subject, see Refs. \cite{Davidson:2000er,Harigaya:2013vwa,Mukaida:2015ria,GA,Passaglia:2021upk,Drees:2021lbm,Drees:2022vvn,Mukaida:2022bbo}} and the energy density of radiation, $\rho_R$, scales as $\rho_R \propto a^{-3/2}$ independent of the spin of the final state products \cite{GKMO2}. This difference in scaling allows for the process of reheating to occur, defined as the moment when $\rho_\phi(\arh) = \rho_R(\arh)$.
Furthermore, if the inflaton primarily decays to vector bosons, this also remains true. However, when the potential minimum is approximated by $\lambda \phi^k$ for $k >2$, the situation changes. The evolution of temperature with respect to the scale factor and the temperature at reheating is highly dependent on whether the decay products are bosons or fermions. In this work, we find that decay to vectors results in a temperature scaling $T \propto a^{-1}$,
which holds regardless of the value of $k$ when $k \ge 4$, and is distinctly different from decays to fermions or scalars \cite{GKMO2}.
Moreover, for $k=2$, the scattering of inflaton condensate to scalars cannot reheat the Universe since the energy density of radiation $\rho_R\propto a^{-4}$ redshifts faster than $\rho_\phi \propto a^{-3}$ \cite{GKMO2}. For $k\ge 4$, the evolution is different, and in this case, two-to-two scatterings of inflatons to scalars can also contribute to reheating. We demonstrate that the same is true for scatterings to vectors as well.

One of our motivations for considering vector decay products stems from models of inflation that are consistent with no-scale supergravity \cite{no-scale}. As the natural framework for a low-energy field theory derived from string theory \cite{Witten}, no-scale supergravity serves as an excellent starting point for the construction of a theory of inflation \cite{building}. In fact, it is straightforward to construct models of inflation \cite{eno6,eno7,enov1,FeKR,egno4,eno9,enov4,building} that lead to Starobinsky inflation \cite{Staro}. However, if there are no direct superpotential couplings between the inflaton and Standard Model fields, there are no decay channels directly to matter scalars or fermions \cite{ekoty,egno4}.\footnote{If the inflaton is associated with a right-handed sneutrino, a superpotential coupling of the inflaton to HL is possible, leading to the production of Higgs bosons and left-handed sleptons \cite{eno8,snu}.} On the other hand, the gauge kinetic function must be field-dependent if gaugino masses are generated at the tree level when supersymmetry is broken.  If the gauge kinetic function depends on the inflaton, it naturally follows that the inflaton could decay into gauge bosons (and gauginos) \cite{ekoty,klor,Terada:2014uia,dlmmo,egno4,building}.
In many models of inflation derived from no-scale supergravity this is the dominant mode for inflaton decay and hence decays to vectors will play a significant role in the reheating process. In this work, we will derive the evolution of the radiation bath when vectors are the primary products of either inflaton decay or scattering.

In addition to the role of vectors in the reheating process, massive vectors may also be produced through scattering, involving either a single graviton exchange or a gravitational portal \cite{Bernal:2018qlk,MO,Bernal:2021kaj,Barman:2021ugy,cmov,Haque:2022kez,cmosv,cmo,Barman:2022qgt,kkmov}. Related mechanisms for the gravitational production of particles during reheating were also considered in \cite{ema,Dimopoulos:2006ms,Graham:2015rva,Garny:2015sjg,Tang:2017hvq,Ema:2019yrd,Chianese:2020yjo,Ahmed:2020fhc,Kolb:2020fwh,Redi:2020ffc,Ling:2021zlj,Ahmed:2021fvt,Haque:2021mab,Aoki:2022dzd,Ahmed:2022tfm,Basso:2022tpd,Haque:2023yra,Kaneta:2022gug,Kolb:2023dzp,Kaneta:2023kfv,Garcia:2023qab,Zhang:2023xcd,RiajulHaque:2023cqe,Ozsoy:2023gnl,Cembranos:2023qph}. In this work, to compute the production of massive spin-1 bosons, we adopt the methods previously used for the production of spin-0 and spin-$\frac12$ particles \cite{MO,cmov}, as well as those for spin-$\frac32$ particles \cite{kkmov}. For related work on the gravitational production of vectors, see \cite{Graham:2015rva,Ema:2019yrd,Kolb:2020fwh,Barman:2021ugy}. Vectors can be produced gravitationally either through the scatterings of SM particles in the newly created thermal bath or directly through inflaton scatterings. We will demonstrate that the latter is by far dominant. Furthermore, we will show that vector bosons produced at the earliest stage of reheating can account for the dark matter if the product of the vector mass and reheating temperature, $m_A \trh \simeq 10^{17}$~GeV$^2$, when $k=2$.  For higher values of $k$, we find  significantly stronger constraints on the vector mass for a given reheating temperature. We compare these results to the production of scalars through gravitational interactions.

The paper is organized as follows: In Section \ref{sec:decays}, we explore the possibility that reheating occurs primarily through the decay of gauge bosons. We compute the evolution of the radiation bath, which originates in the form of vectors, and compute the reheating temperature in terms of $k$, the parameter that determines the shape of the potential about its minimum. In Section \ref{sec:scatterings}, we examine the gravitational production of vectors, considering both scattering within the thermal bath and direct production from the inflaton condensate. Additionally, we discuss two mechanisms that generate vector masses while preserving gauge invariance, namely, the Stueckelberg and Higgs mechanisms. We summarize our results in Section \ref{sec:summ}, and provide some additional details in Appendices~\ref{ap:eff}, \ref{ap:boltz}, and \ref{ap:ampandthermal}.

\section{Inflaton Decay and Annihilation}
\label{sec:decays}

\subsection{The Inflaton Potential}
Reheating is an essential component of any inflationary model, necessitating a coupling between the inflaton sector and the SM. Although gravitational interactions always mediate interactions between the inflaton and the SM, it was shown in \cite{Figueroa:2018twl,Barman:2022qgt} that minimal gravitational couplings alone are insufficient for achieving a radiation-dominated universe with a sufficiently high reheating temperature. Consequently, some form of inflaton decay, or alternatively, an inflaton scattering mechanism, is necessary. In this section, we explore the possibility for non-gravitational decays and scatterings to massive gauge bosons.
We first discuss the motivation for this possibility in the context of no-scale supergravity and subsequently examine the effects of an inflaton-vector coupling (regardless of its origin) on the evolution of the radiation energy density and reheating temperature.

As a phenomenologically acceptable example of an inflaton potential $V(\phi)$, we begin by considering the Starobinsky model of inflation \cite{Staro}
\beq
V(\phi)  \; = \;  \frac34 m_\phi^2 M_P^2 \left(1 - e^{-\sqrt{\frac{2}{3}} \frac{\phi}{M_P}} \right)^2 \, ,
\label{staropot}
\eeq
where $m_\phi$ denotes the inflaton mass, the sole independent parameter in the Starobinsky model. Here, the reduced Planck mass is defined as $M_P = 1/\sqrt{8\pi G_{\rm N}}$. The amplitude of the cosmic microwave background (CMB) anisotropies fixes the inflaton mass to $m_\phi \simeq 3 \times 10^{13}$ GeV. This scalar potential ensures sufficient inflation, and the resulting slow-roll parameters $\epsilon$ and $\eta$ will give a tilt to the CMB anisotropy spectrum. Additionally, it predicts a scalar tilt $n_s$ and the tensor-to-scalar ratio $r$ \cite{eno6,egnov}, both of which are in good agreement with current \textit{Planck} observations \cite{Planck}. Importantly, this potential can be derived from an ultraviolet (UV) perspective in the context of no-scale supergravity. 

\subsection{Motivations from No-Scale Supergravity}

The Starobinsky model can be straightforwardly derived from no-scale supergravity. We define a K\"ahler potential of the following form (in Planck units)
\beq
\label{K0}
K \; = \; -3 \ln \left(T + T^* - \frac{|\varphi|^2}{3} - \frac{|X_i|^2}{3} \right) \, ,
\eeq
where $T$ is associated with the volume modulus, $\varphi$ represents the non-canonical inflaton field,\footnote{The canonical field $\phi$ is related to $\varphi$ though $\phi = \sqrt{3} \tanh^{-1}(\varphi/\sqrt{3})$.} and the $X_i$ are associated with matter fields. The superpotential in the inflationary sector is given by (also expressed in Planck units)
\begin{equation}
W_I(\varphi) = M\left(\frac{1}{2}\varphi^2 - \frac{\lambda}{3\sqrt{3}}\varphi^3 \right)  \, ,
\label{wi}
\end{equation}
leads to the Starobinsky potential once the volume modulus, $T$, is fixed (or stabilized) \cite{eno6}, and the inflaton field $\varphi$ is canonically normalized. Although this superpotential for the Starobinsky potential is not unique \cite{eno7}, various superpotentials that yield the same scalar potential can be related by the underlying SU(2,1)/SU(2)$\times$U(1) no-scale symmetry \cite{enov1}.

If there are no superpotential couplings between $\varphi$ and other SM fields, inflaton decay becomes impossible \cite{ekoty}. As previously mentioned, in this case ($k=2$ for the Starobinsky potential), scatterings do not lead to reheating. An efficient way to introduce 
a coupling of the inflaton to SM fields is through the gauge kinetic function \cite{ekoty,klor,dlmmo,egno4}.
In supergravity, the kinetic terms for gauge fields can be written as
\begin{equation}
  \mathcal{L} \; \supset \; -\frac{1}{4}\left(\Re f_{\alpha\beta} \right)F^\alpha_{\mu\nu}F^{\beta\,\mu\nu}+\frac{i}{4}\left(\Im f_{\alpha\beta}\right)F^\alpha_{\mu\nu}\Tilde{F}^{\beta\,\mu\nu}\label{phiFF} \, ,
\end{equation}
where $F_{\mu\nu}\equiv \partial_\mu A_\nu-\partial_\nu A_\mu$ and $\Tilde{F}_{\mu\nu}\equiv \varepsilon_{\mu\nu\rho\sigma}F^{\rho\sigma}/2$ represent the field strength and its dual, respectively. The inflaton couplings to gauge bosons can arise from the gauge kinetic term if the gauge kinetic function $f_{\alpha\beta}$ depends on the inflaton.

In a supergravity model, an inflaton-dependent modification of the gauge kinetic function leads not only to a coupling of the inflaton to gauge bosons but also to gauginos, which will also be  produced through inflaton decay. The derivative coupling between the gauge boson and the inflaton is given by \eqref{phiFF} and the relevant gaugino-inflaton coupling arises from
\begin{equation}
 \mathcal{L} \; \supset \;
\frac{1}{4} e^{G / 2} \frac{\partial f_{\alpha \beta}^*}{\partial \phi_j^*}\left(G^{-1}\right)_j^k G_k \lambda^\alpha \lambda^\beta+  \textrm{h.c.}
\end{equation}
Here $G \equiv K + \ln W + \ln W^*$ is the K\"ahler function, with $G_i \equiv \partial G/\partial \phi^i$ and $G^j \equiv \partial G/\partial \phi_j^*$, and $\lambda^{\alpha}$ denotes the gaugino fields. By an appropriate choice of the gauge kinetic function
\begin{equation}
    f_{\alpha\beta}\equiv \delta_{\alpha\beta}h(\phi) \, ,
    \label{gaugekin}
\end{equation}
a coupling proportional to $m_\phi \phi \lambda\lambda$ emerges, with the dependence on the inflaton mass originating from the superpotential \eqref{wi}. If we do not associate supersymmetry breaking (at low energies) with the inflationary sector, we can expect that at the minimum, $\phi_{\rm min}$ ($\phi_{\rm min} = 0$ for the Starobinsky potential), $h(\phi_{\rm min}) = 1$, and we restore canonical kinetic terms for the gauge bosons and gauginos.\footnote{To generate gaugino masses at the tree level, it is expected that the function $h$ will contain additional field dependence from the sector responsible for low-energy supersymmetry breaking.}

For the moment, we will focus on the coupling of the inflaton to gauge bosons. Later in this section, we will also explore the implications of decays to gauginos. While we do not specify the function $h(\phi)$, it is clear that  any linear or quadratic term in $\phi$, naturally provides inflaton decay or annihilation channels, respectively, through derivative couplings to the vector bosons. To maintain generality, in the following subsection, we will limit our discussion to couplings of the form\footnote{Gauge boson production during inflation through similar couplings has been discussed in \cite{Barnaby-Namba-Peloso}.}:
\begin{equation}
    \mathcal{L}\supset -\frac{g}{4M_P}\phi F_{\mu\nu}F^{\mu\nu}-\frac{\Tilde{g}}{4M_P}\phi F_{\mu\nu}\Tilde{F}^{\mu\nu}-\frac{\kappa}{4M_P^2}\phi^2 F_{\mu\nu}F^{\mu\nu}-\frac{\Tilde{\kappa}}{4M_P^2}\phi^2 F_{\mu\nu}\Tilde{F}^{\mu\nu} \, ,
    \label{gaugekincouplings}
\end{equation}
where we normalize the coupling constants using the reduced Planck mass, so that $g$, $\Tilde{g}$, $\kappa$, $\Tilde{\kappa}$ are dimensionless.
Note that the coupling of the inflaton to gauge bosons may also induce the non-perturbative production of gauge bosons during inflation, i.e., superhorizon modes, which is on the other hand strongly dependent on the form of $h(\phi)$ for $\phi\gg M_P$.
We here assume that the couplings (\ref{gaugekincouplings}) are valid only after the end of inflation and will focus on the gauge boson production during reheating.

\subsection{Reheating Through Decays to Vector Bosons}
\label{Sec:decaytovectors}
The most direct reheating process involves the decay of the inflaton field. However, computing particle production through the decay of a scalar condensate such as the inflaton can differ from calculating the decay width of a quantum field. For oscillations about a quadratic potential, the results are identical. In more general cases, when the potential can be expanded around its minimum as
\begin{equation}
    V(\phi)=\lambda M_P^4 \left( \frac{\phi}{M_P}\right)^k \,, \quad \phi\ll M_P \, ,
    \label{kexp}
\end{equation} 
the derived rates depend on an effective coupling, averaged over a single oscillation, which, in turn, depends on the shape of the potential and hence the parameter $k$. In general, these effective couplings must be computed numerically \cite{Shtanov:1994ce,Ichikawa:2008ne,GKMO2}, and details of this computation are given in Appendix \ref{ap:eff}.

The Starobinsky potential, when expanded about the minimum, gives $k=2$ and does not easily generalize to larger values of $k$. However, the related $\alpha$-attractor T-models~\cite{Kallosh:2013hoa} are also phenomenologically viable \cite{egnov}. The scalar potential for these models can be expressed as 
\beq
\label{eq:Vattractor}
V(\phi) \;=\; \lambda M_P^4 \left[ \sqrt{6} \tanh\left(\frac{\phi}{\sqrt{6} M_P }\right)\right]^k\,,
\eeq
which reduces to Eq.~(\ref{kexp}) when expanded about the minimum at $\phi = 0$. We note that this class of potentials can also be generated from no-scale supergravity models by choosing the following superpotential form \cite{GKMO1}
\beq
W \; = \; 2^{\frac{k}{4}+1}\sqrt{\lambda}  \left( \frac{\phi^{\frac{k}{2}+1}}{k+2} - \frac{\phi^{\frac{k}{2}+3}}{3(k+6)} \right)\, .
\eeq
As in the Starobinsky model, the lone parameter $\lambda$ is determined from the normalization of the CMB anisotropies \cite{Planck}. The normalization of the potential for different values of $k$ is given by~\cite{GKMO2}
\begin{equation}
    \label{eq:normlambda}
    \lambda \; \simeq \; \frac{18\pi^2 A_{S*}}{6^{k/2} N_*^2} \, ,
\end{equation}
where $N_*$ is the number of e-folds of inflation and $A_{S*} \simeq 2.1 \times 10^{-9}$ is the amplitude of the curvature power spectrum. For this potential with $k=2$, we find $\lambda = 2.05 \times 10^{-11}$ and $m_\phi = \sqrt{2\lambda} M_P = 1.5 \times 10^{13}$ GeV for $N_* = 55$ e-folds.

From Eq.~(\ref{gaugekincouplings}), the inflaton decay rate to vector bosons can be computed as
\beq
 \Gamma_{\phi\rightarrow A_\mu A_\mu} \; = \;
 \frac{\alpha^2 m_\phi^3}{M_P^2}\,,
 \label{Eq:gammadecay}
\eeq
where $\alpha^2 = (g_{\rm eff}^2 + \tilde g_{\rm eff}^2)/(64\pi)$
and the effective couplings $g_{\rm eff}$ and $\tilde g_{\rm eff}$ arising from the Lagrangian~(\ref{gaugekincouplings}) are given by Eqs.~(\ref{Eq:geff}) and~(\ref{Eq:gtildeeff}), respectively. The Lagrangian coupling $g$ (${\tilde g}$) is defined by Eq.~(\ref{Eq:geff}) (Eq.~(\ref{kexp})) 
in Appendix \ref{ap:eff}. Note that the couplings to $FF$ and $F\Tilde{F}$ do not interfere due to the antisymmetry of the Levi-Civita tensor. The inflaton mass
is given by
\beq m^2_\phi=\frac{\partial^2 V(\phi)}{\partial \phi^2}=k(k-1)\lambda^\frac{2}{k}M_P^{\frac{8-2k}{k}}\rho_\phi^{\frac{k-2}{k}}\,.
\eeq
For the last equality, we use $\rho_\phi = V(\phi_0)$, given by Eq.~(\ref{kexp}), where $\phi_0$ is the time-dependent amplitude of the oscillation of $\phi$, as discussed in more detail in Appendix \ref{ap:boltz}. 

Following the procedure described in \cite{GKMO2}, it is useful to rewrite the decay rate in terms of $\rho_\phi$
so that
\beq
\Gamma_{\phi\rightarrow A_\mu A_\mu}=\gamma_\phi
\left(\frac{\rho_\phi}{M_P^4}\right)^l
\eeq
with
\beq
l=\frac{3}{2}-\frac{3}{k}\,,
\eeq
and
\beq
\gamma_\phi= \alpha^2 \left[k(k-1)\right]^{\frac{3}{2}}\lambda^{\frac{3}{k}}M_P\,.
\eeq
The dependence of the decay rate on the energy density in this case differs from that of decays to fermions or scalars. 
In the case of decays to fermions, $\phi\rightarrow f \bar f$, the power is given by $l=\frac{1}{2}-\frac{1}{k}$, whereas for decays to scalar bosons $\phi \rightarrow bb$, $l=\frac{1}{k}-\frac{1}{2}$ \cite{GKMO2}. As expected, they follow the redshift behavior $\Gamma_\phi\propto m_\phi$
for the fermion final state, and $\Gamma_\phi\propto 1/m_\phi$ for boson final states. In the case we are considering here, the form of the couplings (\ref{gaugekincouplings}) imposed by gauge invariance implies a width which undergoes a much 
more pronounced redshift, with $\Gamma_\phi \propto m_\phi^3$. This dependence is very important, because it indicates that the behaviour of vector reheating is closer to that of fermions (a width that decreases with time) than to scalar boson decay (a width that increases with time). Furthermore, since the decay rate decreases over time, the energy transfer is most effective at the beginning of the process.

The decay rate determines the evolution of the radiation density  through the coupled equations of motion:
\bea
\dot{\rho}_{\phi} + 3H(1+w_{\phi})\rho_{\phi} \;\simeq\; -\Gamma_{\phi}(1+w_{\phi})\rho_{\phi} \, ,
\label{rhoeom1} \\
\dot{\rho}_{R} + 4H \rho_{R} \;\simeq\; (1+w_{\phi})\Gamma_{\phi}(t)\rho_{\phi}\, ,
\label{aeom}
\eea
where $H=\frac{\dot a}{a}$ is the Hubble parameter, $a$ is the cosmological scale factor, and the equation of state parameter $w_{\phi} = \frac{P_\phi}{\rho_\phi}$ is given by
\beq\label{eq:wk}
w_{\phi} \;=\; \frac{k-2}{k+2}\,.
\eeq
Using $\frac{d}{dt} = a H \frac{d}{da}$, we can integrate Eq.~(\ref{rhoeom1}) to obtain 
\beq
\rho_\phi(a) = \rho_{\rm end} \left(\frac{a}{a_{\rm end}} \right)^{-\frac{6k}{k+2}} \, ,
\label{Eq:rhophi}
\eeq
valid when $\gamma_\phi \ll H$. 
Here $\rho_{\rm end} = \rho_\phi(a_{\rm end})$, where $a_{\rm end}$ is the value of the scale factor when $\phi(a_{\rm end}) = \phi_{\rm end}$ and $\phi_{\rm end}$ corresponds to the value of $\phi$ when the inflationary expansion ends, namely when $\ddot{a} = 0$ which implies that $\rho_{\rm end} = \frac32 V(\phi_{\rm end})$. An approximate solution for the inflaton field value at the end of inflation as a function of $k$ is given by~\cite{GKMO2}
\begin{equation}
    \label{eq:phiendapprox}
    \phi_{\rm end} \; = \; \sqrt{\frac{3}{8}} M_P \ln \left[\frac{1}{2} + \frac{k}{3} \left( k + \sqrt{k^2 + 3 }\right) \right] \, .
\end{equation}
For $k=2$, we find $\rho_{\rm end}^{1/4} = 5.2 \times 10^{15}$~GeV, and it slightly decreases with increasing $k$ \cite{GKMO2,cmov}.

Using the solution for $\rho_\phi$, we integrate Eq.~(\ref{aeom}) and obtain
\begin{equation}
\rho_R \; = \; \frac{2 k}{k+8-6 k l} \frac{\gamma_\phi}{H_{\mathrm{end}}} \frac{\rho_{\mathrm{end}}^{l+1}}{M_P^{4 l}}\left(\frac{a_{\mathrm{end}}}{a}\right)^4\left[\left(\frac{a}{a_{\mathrm{end}}}\right)^{\frac{k+8-6 k l}{k+2}}-1\right] \,,
\label{rhoR}
\end{equation}
so at late times, when $a\gg a_\text{end}$, and for $8+k-6kl=26-8k>0 $, we obtain
\begin{equation}
T=\left(\frac{30\rho_R}{g_\rho\pi^2}\right)^\frac{1}{4}\propto a^{-\frac{3k+6kl}{4k +8}}=a^{\frac{9-6k}{4+2k}} \, ,
\label{tvsa}
\end{equation}
where $g_\rho = 915/4$ is the number of relativistic degrees of freedom which we take from the minimal supersymmetric standard model (MSSM). 
When $k=2$ and $l=0$, $T \propto a^{-\frac{3}{8}}$ 
which is the same behavior as other decay channels, $\phi\rightarrow f \bar{f}$ and  $\phi\rightarrow bb$ because the width $\Gamma_\phi$
is constant for a quadratic potential. However, for $k\geq 4$, we have $8+k-6kl<0$, and  the radiation density in Eq.~\eqref{rhoR} is dominated by the second term,  and the temperature evolves as $\rho_R \propto a^{-4}$. 

This result illustrates what we discussed above: reheating primarily occurs at the very beginning of the process, subsequently, the density then follows the classical isentropic redshift law, where the decay rate $\propto \rho_\phi^{\frac{3}{2}-\frac{3}{k}}$ redshifts faster than the expansion rate $H\propto \rho_\phi^{\frac{1}{2}}$. In addition, the inflaton energy density from Eq.~(\ref{Eq:rhophi}) scales $\rho_\phi\propto a^{-\frac{6k}{k+2}}$, which gives $a^{-3}$, $a^{-4}$, $a^{-\frac{9}{2}}$ for $k=2,4,6$, respectively. Consequently, reheating is not possible for $k=4$ since $\rho_\phi$ and $\rho_R$ scale in the same way. However, reheating is possible for $k=2$ and $k>4$. Nonetheless, as we discuss below, $k > 10$ is necessary for reheating temperatures higher than $\trh > 100$ GeV. 

This behavior can be seen from the analytic expressions for the reheating temperature. Since we define the reheat temperature when 
\beq
\rho_R(T_{\rm RH}) = \rho_\phi(T_{\rm RH}) \, ,
\eeq
we can determine $\arh/\aend$ to find $\rho_{\rm RH} = \rho_R(\arh)$, which for $8+k-6kl >0$ gives 
\begin{align}
T_{\rm RH} \;&=\; \left(\frac{30}{g_\rho \pi^2} \right)^{\frac{1}{4}}
\left[\frac{2k}{k+8-6kl} 
\frac{\sqrt{3}\gamma_\phi}{M_P^{4l-1}} \right]^{\frac{1}{2-4l}} \\
& = \left(\frac{30}{g_\rho \pi^2} \right)^{\frac{1}{4}} \left( 8 \sqrt{6} \lambda^\frac32/10 \right)^\frac12 \alpha M_P \qquad k=2, l=0 \, .
\end{align}
For $8+k-6kl < 0$ and $k> 4$,
\begin{align}
T_{\rm RH} \;&=\; \left(\frac{30}{g_\rho \pi^2} \right)^{\frac{1}{4}}
\left[\frac{2k}{6kl-k-8} 
\frac{\sqrt{3}\gamma_\phi}{M_P^{4l-1}} \rho_{\rm end}^{\frac{6kl-k-8}{6k}}\right]^{\frac{3k}{4k-16}} \\
& = \left(\frac{30}{g_\rho \pi^2} \right)^{\frac{1}{4}} \left[\frac{2\sqrt{3}k}{8k-26} 
\alpha^2 (k(k-1))^\frac32 \lambda^\frac3k \right]^{\frac{3k}{4k-16}}
\left(\frac{\rho_{\rm end}}{M_P^4} \right)^{\frac{8k-26}{8k-32}} M_P\, ,
\label{dkg4}
\end{align}
which is clearly seen to be proportional to $\rho_{\rm end}$ for large $k$.

In the absence of a term linear in $\phi$ in $h(\phi)$, a quadratic term may also produce vector bosons through 
annihilation, $\phi\phi \rightarrow A_\mu A_\mu $, driven by the last two terms of Eq.~\eqref{gaugekincouplings}. These two couplings will give rise to the same dependence of the rate on $\rho_\phi$ and $m_\phi$. Defining $\beta^2\equiv (\kappa_{\rm eff}^2 +\Tilde\kappa_{\rm eff}^2)/(4\pi)$,
where $\kappa_{\rm eff}$ and $\Tilde \kappa_{\rm eff}$ are given 
as a function of $\kappa$ and 
$\Tilde \kappa$ by Eqs.~(\ref{Eq:kappaeff}) and (\ref{Eq:kappatildeeff}), respectively, we have 
\begin{equation}
 \Gamma_{\phi \phi\rightarrow A_\mu A_\mu}=\frac{\beta^2 \rho_\phi }{M_P^4}m_\phi \,,
 \label{Eq:scattering}
\end{equation}
from which we can extract
\begin{equation}
    \begin{aligned}
        &l=\frac{3}{2}-\frac{1}{k}\,,\\
        &\gamma_\phi=\beta^2 \sqrt{k(k-1)}\lambda^{\frac{1}{k}}M_P\,.
    \end{aligned}
\end{equation}

As  $8+k-6kl=14-8k$ is always negative when $k\geq 2$, the second term in the brackets in Eq.~(\ref{rhoR}) dominates and the temperature decreases as $T\propto a^{-1}$ at late times. 
This is because $\Gamma_\phi \propto a^{-\frac{9k-6}{k+2}}$ whereas $H \propto a^{-\frac{3k}{k+2}}$. For $k\geq 2$, 
the scattering rate decreases faster than the Hubble rate, and reheating occurs at the 
very end of inflation. Very quickly, the rate of energy transfer from the 
condensate to the thermal bath is unable to compensate for the 
expansion rate. However, as $\rho_\phi\propto a^{-\frac{6k}{k+2}}$ and $\rho_R\propto a^{-4}$, 
reheating can only take place if $k\geq 6$.
The reheating temperature in this case is given by
\begin{align}
T_{\rm RH} \;&=\;  \left(\frac{30}{g_\rho \pi^2} \right)^{\frac{1}{4}} \left[\frac{2\sqrt{3}k}{8k-14} 
\beta^2 (k(k-1))^\frac12 \lambda^\frac1k \right]^{\frac{3k}{4k-16}}
\left(\frac{\rho_{\rm end}}{M_P^4} \right)^{\frac{4k-7}{4k-16}} M_P\, .
\label{skg4}
\end{align}
This situation is very different from scalar boson 
scattering \cite{GKMO2}. 
In this case, the radiation density evolves as $\rho_R \propto a^{-\frac{18}{k+2}}$, which means that reheating 
is impossible for $k=2$, but feasible 
for $k\geq 4$. Moreover, in the case 
of scalar boson scattering, for $k \geq 4$ reheating is most efficient at the end of the process, not at the beginning. The behavior of the temperature during reheating is given in Table \ref{Tab:table}
for decays into fermions, scalars, and vectors as well as scattering to scalars and vectors. 

\begin{table*}[!ht]
\centering
\bgroup
\def\arraystretch{1.5}
\begin{tabular}{|c||c|c|c|c|}
\hline 
channel & generic & $k=2$ & $k=4$ & $k=6$ 
 \\ 
 \hline \hline
$\phi\rightarrow f \bar f$ & $T\propto a^{-\frac{3k-3}{2 k + 4}}$ & $T \propto a^{-3/8}$ & $T \propto a^{-3/4}$ & $T \propto a^{-15/16}$ 
\\
 \hline 
$\phi\rightarrow bb$ & $T\propto a^{-\frac{3}{2k+4}}$ & $T \propto a^{-3/8}$ & $T \propto a^{-1/4}$ & $T \propto a^{-3/16}$ 
\\
\hline 
$\phi\rightarrow A A$ & $T\propto a^{-\frac{6k-9}{2k+4}}$ & $T \propto a^{-3/8}$ & $T \propto a^{-1}$ & $T \propto a^{-1}$ 
\\
 \hline 
$\phi \phi \rightarrow b b$ & $T\propto a^{- \frac{9}{2 k + 4}}$ & $T \propto a^{-1} $ & $T \propto a^{-3/4}$ & $T \propto a^{-9/16}$
\\
\hline 
$\phi \phi \rightarrow A A$ & $T\propto a^{- \frac{6k-3}{2 k + 4}}$ & $T \propto a^{-1} $ & $T \propto a^{-1}$ & $T \propto a^{-1}$
\\
\hline
\end{tabular}
\egroup
\caption{Dependence of the temperature $T$ as function of the scale factor $a$ for the different cases that we analyze in this work. The `generic' result assumes the validity of Eq.~(\ref{tvsa}). This is not true when $8+k-6kl <0$, in which case the scaling is simply $T \propto a^{-1}$.}
\label{Tab:table}
\end{table*}

\begin{figure}[!ht]
  \centering
\includegraphics[width=\textwidth]{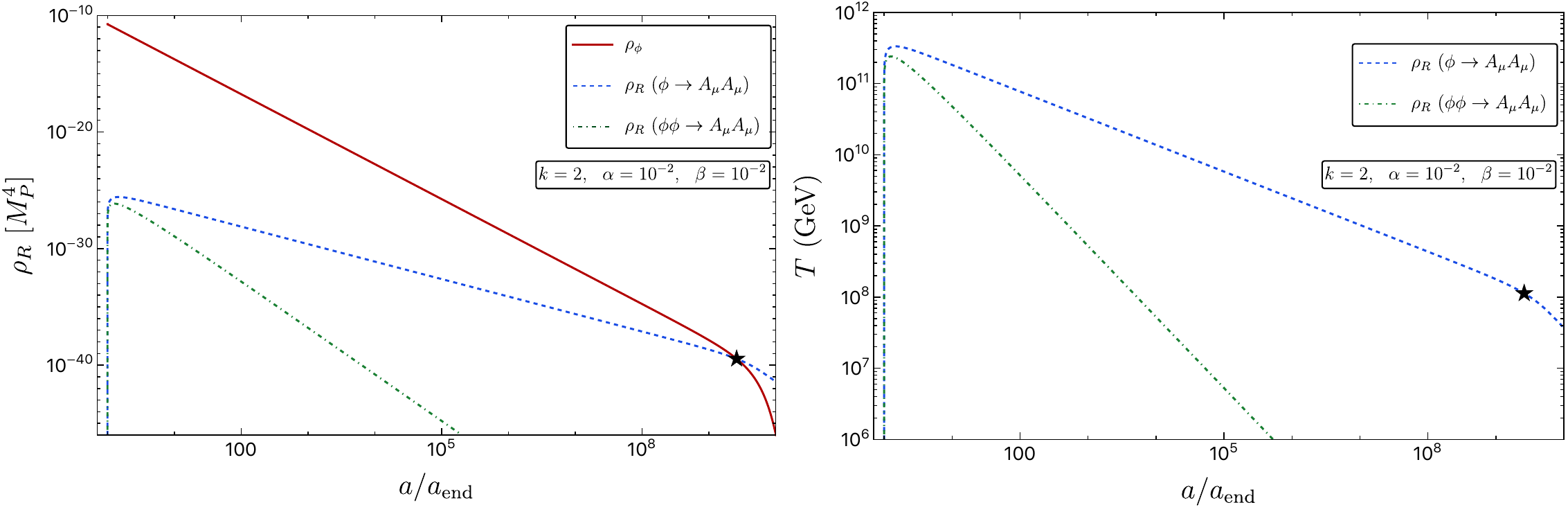}
  \caption{\em \small {
The evolution of the energy density of inflaton, $\rho_\phi$ (solid red), and radiation, $\rho_R$  (blue dashed for decay and green dot-dashed for scattering) as a function of the scale factor $a/a_{\rm end}$ for $k=2$, $\alpha=10^{-2}$ and $\beta=10^{-2}$ (left panels). The evolution of the temperature of the radiation is shown in the right panels. The star indicates the moment when $\rho_R(T_{\rm RH}) = \rho_\phi(T_{\rm RH})$.
} 
}
  \label{Fig:rhok2}
\end{figure}

\begin{figure}[!ht]
  \centering
\includegraphics[width=\textwidth]{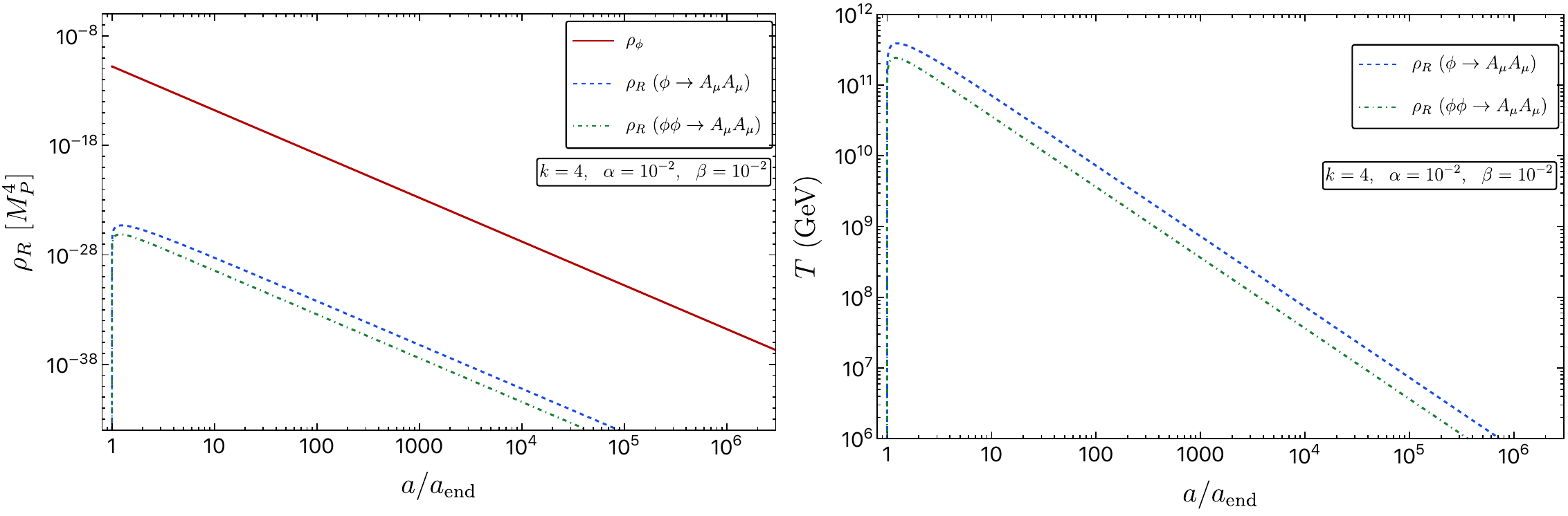}
  \caption{\em \small As in Fig.~\ref{Fig:rhok2}, for $k=4$.
}
  \label{Fig:rhok4}
\end{figure}

\begin{figure}[!ht]
  \centering
\includegraphics[width=\textwidth]{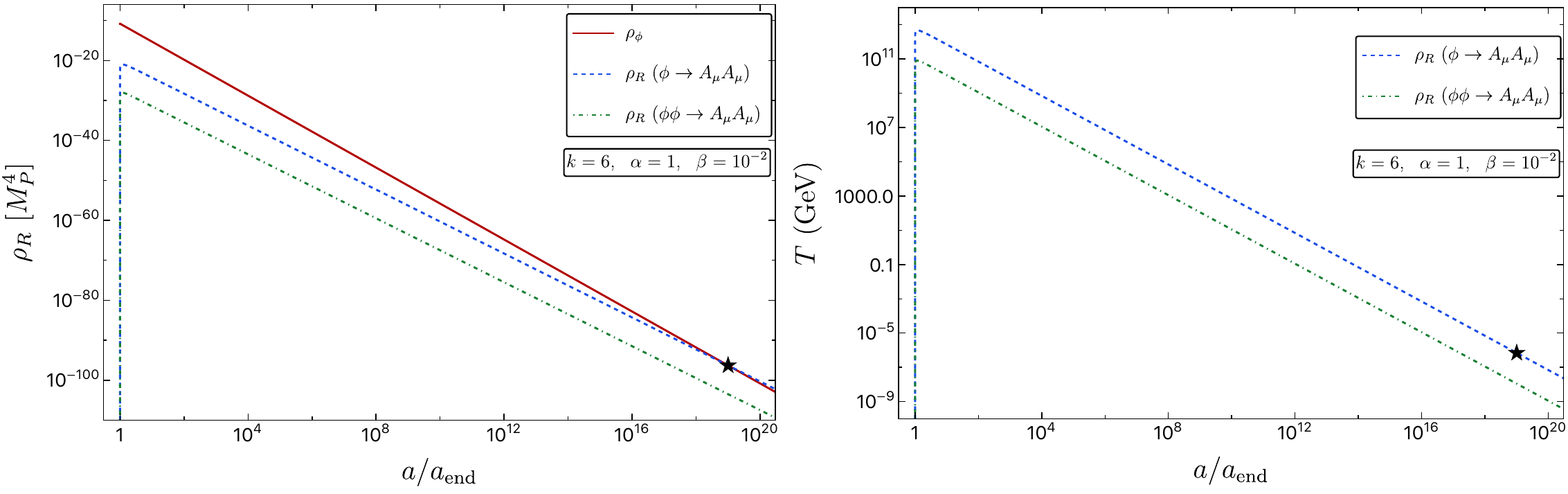}
  \caption{\em \small As in Fig.~\ref{Fig:rhok2}, for $k=6$, $\alpha = 1$ and $\beta = 10^{-2}$.  
}
  \label{Fig:rhok6}
\end{figure}

We show in Figs.~\ref{Fig:rhok2}-\ref{Fig:rhok6} the evolution of $\rho_\phi$ and $\rho_R$ for $k=2,4$, and 6 in the case of inflaton decay and scattering for $\alpha=10^{-2}$ ($\alpha = 1$ for $k = 6$) and $\beta=10^{-2}$, respectively, in the left panels. The temperature of the radiation bath is shown in the right panels and the reheating temperature is indicated by a star.  For $k=2$, shown in Fig.~\ref{Fig:rhok2}, we see that reheating through the decay is possible, and quite efficient, giving a reheating temperature of $\trh\sim 10^{9}$ GeV for the adopted value of $\alpha$. On the other hand, as we predicted, reheating is not possible via scattering in 
this case, since $\rho_R\propto a^{-4}$ dilutes more 
rapidly than $\rho_\phi\propto a^{-3}$. 
For $k=4$, Fig.~\ref{Fig:rhok4} shows that neither inflaton decay nor scattering to vectors is efficient enough to achieve reheating. Indeed, in both cases, the
inflaton density and the energy density of the radiation decrease as $a^{-4}$. However, as demonstrated in Fig.~\ref{Fig:rhok6}, reheating is possible for both decays and scatterings when $k\geq 6$. For this value of $k$, with $\alpha =1$ and $\beta = 10^{-2}$, we see that reheating from decays occurs very late resulting in a very low reheating temperature, as seen in the right panel. 
Reheating from scatterings also occurs but for this value of $\beta$, it occurs even later, beyond the range shown in the plot. 
Though it is hard to see, the slopes for $\rho_R$ from scattering is shallower than $\rho_\phi$. 

\begin{figure}[!ht]
  \centering
\includegraphics[width=0.7\textwidth]{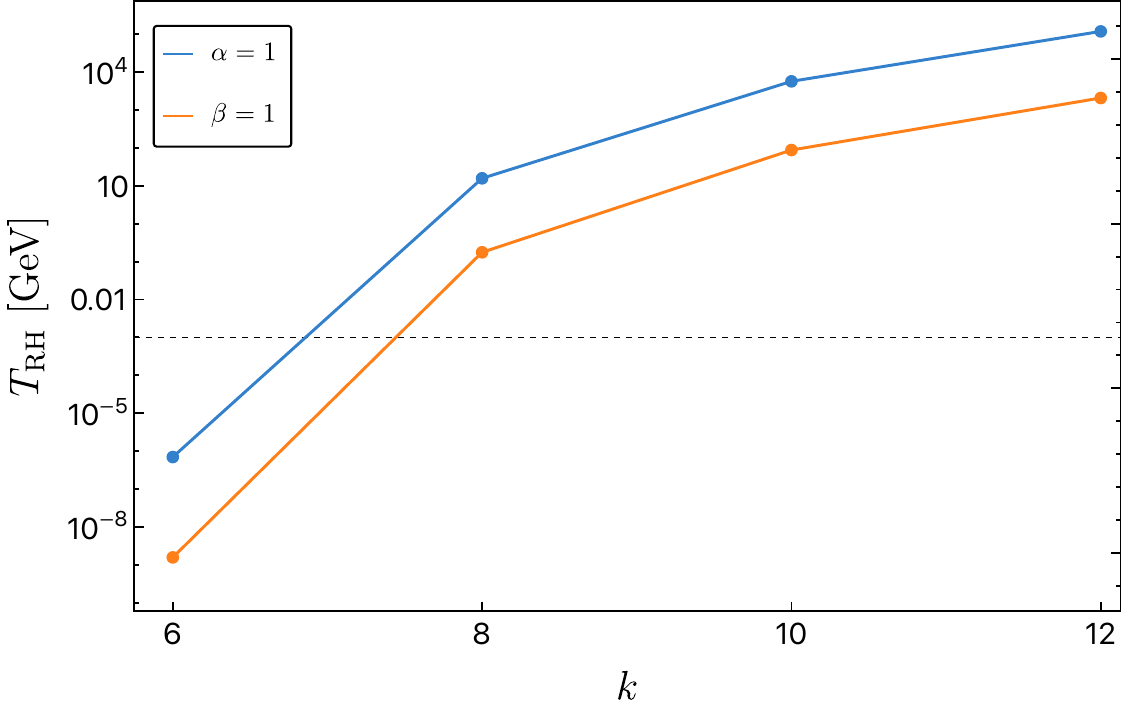}
  \caption{\em \small The reheating temperature defined when $\rho_\phi = \rho_R$ as a function $k$ for $k\ge 6$ in the case of decays (solid blue) with $\alpha = 1$ and scatterings (solid orange) with $\beta  = 1$. 
}
  \label{Fig:trehvsk}
\end{figure}

As one can infer from the $k$ dependence in Eqs.~(\ref{dkg4}) and (\ref{skg4}), the reheating temperature for $k>4$ increases with increasing $k$. This is demonstrated in Fig.~\ref{Fig:trehvsk}, where we show the resulting reheating temperature as a function of $k$. Here we have assumed $\alpha = 1$ and $\beta = 1$ for decays and scatterings, respectively. As already noted,
the reheating temperature is too small for $k=6$, but is considerably higher when $k \ge 8$. 

The analysis presented above is perturbative, and in scenarios when reheating is significantly delayed (e.g. when $k=6$), non-perturbative effects such as the fragmentation of the condensate become important \cite{Amin:2011hj,Lozanov:2016hid,Lozanov:2017hjm,Antusch:2021aiw,Garcia:2023eol,Garcia:2023dyf}. For the above perturbative analysis to hold, reheating must occur before the fragmentation of the condensate.
The fragmentation of the condensate for the T-models considered here has been recently examined for $k \ge 6$ \cite{Garcia:2023dyf}. As one might expect, the relative value of the scale factor, $a_{\rm frag}/\aend$ increases substantially with $k$, whereas $\arh/\aend$ decreases with $k$.
We find that the $\arh \sim a_{\rm frag}$ for $k=10$ and $\alpha = 1$. Note that $\arh/\aend \propto \alpha^{-(k+2)/(k-4)}$ ($a_{\rm frag}/\aend$ is independent of $\alpha$). For $k = 12$, reheating occurs before fragmentation for $\alpha \gtrsim \mathcal{O}(0.1)$,
and the limit on $\alpha$ weakens at higher $k$. Thus, for values of $k$ sufficiently large to provide a reasonable reheating temperature, the effects of fragmentation can be safely ignored. Although similar constraints could be imposed on $\beta$, as evident from Fig.~\ref{Fig:trehvsk}, achieving the same reheating temperature would require larger values of $\beta$, and the fragmentation limits discussed above would also correspond to increased values of $\beta$.

\subsection{Production of gauginos}

\begin{figure}[!ht]
\centering
\includegraphics[width=6.5in]{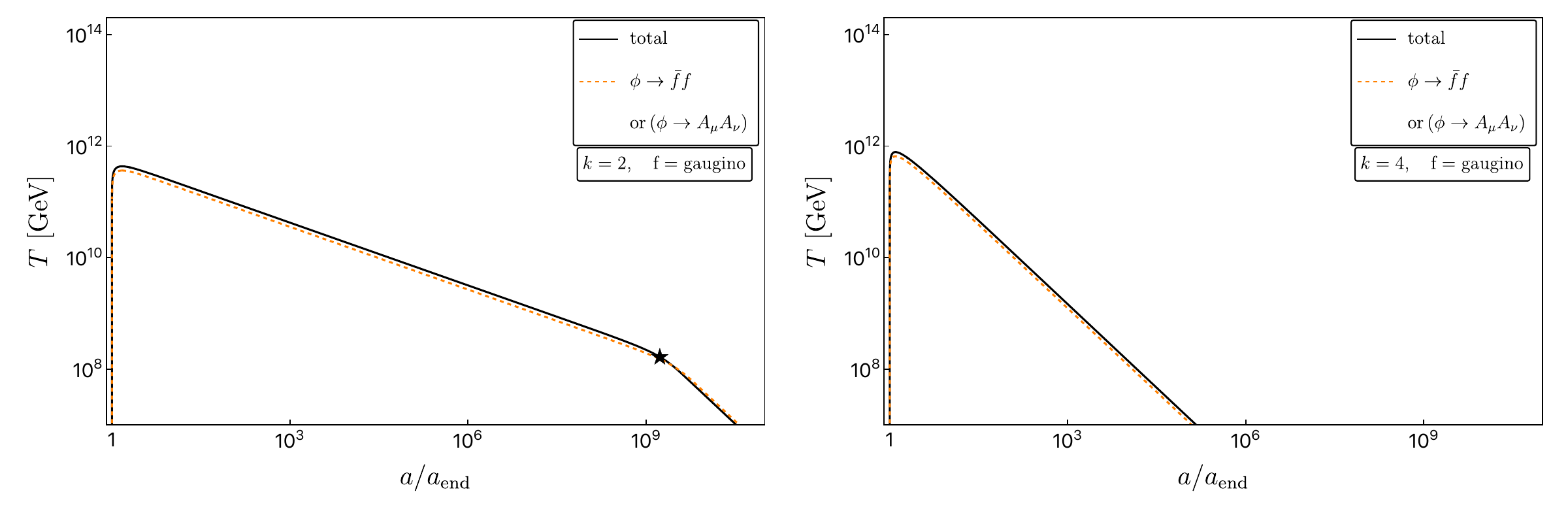}
\caption{\em \small The evolution of the temperature as a function of $a/\aend$ for decays to either gauge bosons or gauginos (dashed lines) and the sum when both are included (solid lines)  for $k=2$ (left) and $k=4$ (right). Here we have chosen 
$\alpha=10^{-2}$ and $g_\rho=915/4$. The star indicates the temperature and scale factor at reheating.
}
\label{Fig:gauginoplot}
\end{figure}

Before concluding this section, we recall that in a supersymmetric model, generally the decays of the inflaton to gauge bosons will be accompanied by decays to gauginos (so long as the channel is kinematically accessible).  
The results discussed above and in \cite{GKMO2} assume that there is a dominant decay channel with definite spin. 
However, 
it was shown in \cite{egno4} that the partial width to gauginos is the same as the partial width to gauge bosons  \begin{equation}
\Gamma_{\phi\rightarrow A_\mu A_\mu}=\Gamma_{\phi\rightarrow \lambda\lambda}={y^\prime}^2 \frac{m_\phi^3}{M_P^2} \, ,
\end{equation}
where $y^\prime$ is a dimensionless factor  encoding the number of final states and the form of the gauge kinetic function. The above equality implies that the gauge boson and the gaugino result in exactly the same evolution of the reheating temperature, namely, $T\propto a^{-\frac{3}{8}}$ for $k=2$ and $T\propto a^{-1}$ for $k\geq4$. Compared to the case with only gauge bosons in the final state, the  effect of a gauge boson-gaugino mixture in the  final states amounts simply to doubling the decay rate in the $\rho_R$ and $\rho_\phi$ evolution equations. In Fig.~\ref{Fig:gauginoplot}, we show the numerical solution of the reheating temperature when decays to gauge bosons and gauginos are included. As one can see, there are no changes in the evolutionary slopes.  

\begin{figure}[!ht]
\centering
\includegraphics[width=6.5in]{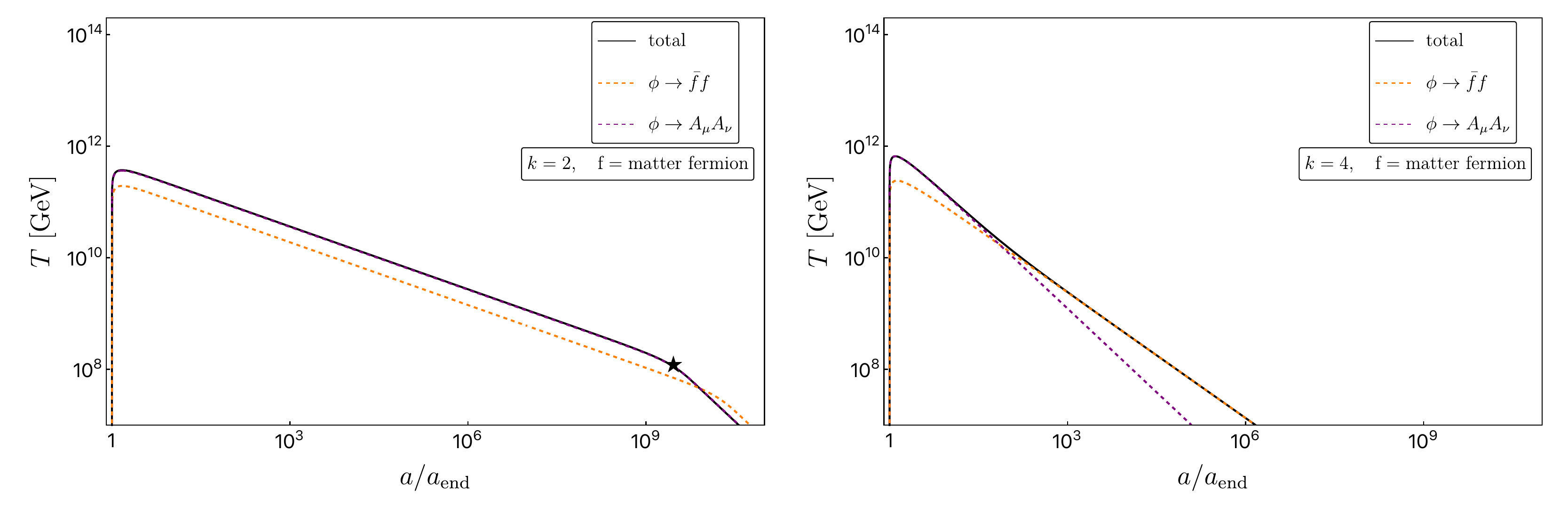}
\caption{\em \small 
As in Fig.~\ref{Fig:gauginoplot}
comparing the evolution of the temperature for decays to a matter fermion (yellow dashed) and to gauge bosons (purple dashed) and the total (solid black) for $k=2$ (left) and $k=4$ (right). We have taken
$y=10^{-7}$, $\alpha=10^{-2}$, and $g_\rho=915/4$.
}
\label{Fig:mfermion}
\end{figure}

It is interesting to note that for a generic Yukawa coupling $y\phi \bar{f}f$ where $y$ is independent of the inflaton mass, the inflaton decay rate differs from $\Gamma_{\phi\rightarrow \lambda\lambda}$. The former is given by \cite{GKMO2}
\begin{eqnarray}
   \Gamma_{\phi \rightarrow \bar f f}=\frac{y^2}{8\pi}m_\phi \, .
\end{eqnarray}
In this case, the  temperature scales as $T\propto a^{-3/8}, a^{-3/4}, a^{-15/16}$, respectively, for $k=2,4,6$, as seen in the Table, and $T\propto a^{-1}$ for $k\geq8 $. As a result, the qualitative behavior of the temperature is different for a gauge boson-gaugino mixture and for a gauge boson-matter fermion mixture in the range $4\leq k\leq 6$. In Fig.~\ref{Fig:mfermion}, we show the numerical solution for $T$ in the case of gauge boson-matter fermion mixing for $k=2$ (left) and $k=4$ (right). As one can see, for $k=2$, since all decay channels lead to the same scaling ($T\propto a^{-3/8}$),
the lines are parallel, though for $k=4$, they are not.

Note that the above discussion neglects the effects of the kinematic suppression due to the effective masses of the fermions and inflaton \cite{GKMO2}.
When there is a non-zero field value for the inflaton, the coupling leading to the decay of the condensate also provides an effective mass to fermions (and bosons, though not to gauge bosons).\footnote{Gaugino masses are also generated when $h_\phi$ is non-vanishing and supersymmetry is broken, which is typical when the inflaton is not yet settled in its potential minimum.} This effect leads to a suppression in the decay rate to fermions proportional to $\mathcal{R}^{-1/2}$ with $\mathcal{R} \propto m_{\rm eff}^2/m_\phi^2$. For example, for the generic Yukawa coupling, $m_{\rm eff} = y \phi$, and $m_\phi^2 = k(k-2) \lambda \phi^{(k-2)} M_P^{(4-k)}$, and $\mathcal{R} \propto (\phi/M_P)^{(4-k)}$ and can be quite large for $k \ge 6$. Thus, for large $k$, we expect that our results for gauge boson final states are unaffected by the inclusion of fermions.

\section{Gravitational production}
\label{sec:scatterings}
\begin{figure}[!ht]
\centering
\includegraphics[width=3.5in]{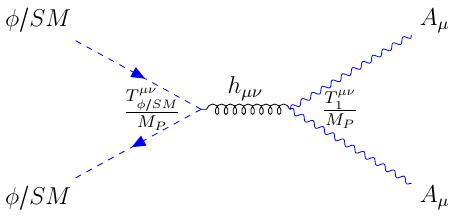}
\caption{\em \small Feynman diagram for the production of massive vector particles through the gravitational scattering of the inflaton condensate or the Standard Model particle bath. 
}
\label{Fig:feynman}
\end{figure}

In the previous section, we have considered the possible role for vector bosons in the reheating process. That is, given a coupling between the inflaton and some vector boson produced either by decay or scattering, the evolution of the radiation bath differs from the cases where the inflaton is coupled to fermions or scalars. Gauge invariance dictates a derivative coupling leading to a Planck mass suppression (in the case motivated by supergravity in Eq.~(\ref{gaugekincouplings})). A non-suppressed coupling to vectors would lead to an evolution identical to that in the case of scalars. However, within the Standard Model, there is an unavoidable coupling between a massive vector and the inflaton, mediated by gravity. Indeed, the scattering through the exchange of an $s$-channel graviton, as shown in Fig.~\ref{Fig:feynman}, is a process that {\it must} be present, and therefore provides a {\it minimal} (unavoidable) production of vector bosons (and potentially a dark matter candidate) during the reheating process. In this section, we calculate the abundance of vectors produced solely through gravitational interactions.

\subsection{The amplitude}

To compute the gravitational production rate of vector bosons, we begin by first expanding the space-time metric around a flat Minkowski background. Using the relation $g_{\mu \nu} \simeq \eta_{\mu \nu} + \frac{2h_{\mu \nu}}{M_P}$, with $h_{\mu \nu}$ being the canonically normalized graviton field, the interaction Lagrangian can be expressed as~\cite{cmov,cmosv,hol}
\begin{equation}
    \sqrt{-g} \mathcal{L}_{\rm int} \; = \; -\frac{1}{M_P} h_{\mu\nu}\left(T^{\mu \nu}_{\rm SM} + T^{\mu \nu}_{\phi} + T^{\mu \nu}_{A_{\mu}} \right) \, ,
\end{equation}
where $T^{\mu \nu}_{\rm SM}$, $T^{\mu \nu}_{\phi}$, and $T^{\mu \nu}_{A_{\mu}}$ represent the energy-momentum tensors for the Standard Model particles, the inflaton, and the massive vector respectively. We will consider a  massive vector field described by the Proca Lagrangian:\footnote{We discuss the generation of a mass term for the vector field through the Stueckelberg and Higgs mechanisms in Section~{\ref{stueck}}.}
\begin{equation}
\begin{aligned}
     \mathcal{L} \; = \; -\frac{1}{4}F^{\mu\nu}F_{\mu\nu}+\frac{1}{2}m_A^2 A^\mu A_\mu \, ,
\end{aligned}
\label{procaL}\end{equation}
and the canonical energy-momentum tensors for the massive fields with spins $i=0,1/2, 1$ are given by
\bea
T^{\mu \nu}_{0} &=&
\partial^\mu S \partial^\nu S-
\eta^{\mu \nu}
\left[
\frac{1}{2}\partial^\alpha S \partial_\alpha S-V(S)\right] \, ,
\label{Eq:tensors}
\\
T^{\mu \nu}_{1/2} &=&
\frac{i}{4}
\left[
\bar \chi \gamma^\mu \overset{\leftrightarrow}{\partial^\nu} \chi
+\bar \chi \gamma^\nu \overset{\leftrightarrow}{\partial^\mu} \chi \right] -\eta^{\mu \nu}\left[\frac{i}{2}
\bar \chi \gamma^\alpha \overset{\leftrightarrow}{\partial_\alpha} \chi
-m_\chi \bar \chi \chi\right] \, , \\
\label{Eq:tensorf}
T_{1}^{\mu \nu} &= &  -\frac{1}{2} \left[F_{\alpha}{}^{\mu} F^{\alpha \nu}+F_{\alpha}{}^{\nu} F^{\alpha \mu}  - \frac{1}{2} \eta^{\mu \nu} F^{\alpha \beta} F_{\alpha \beta} + \eta^{\mu \nu} m_{A}^2 A_{\alpha} A^{\alpha} -2 m_{A}^2 A^{\mu} A^{\nu}\right] \, ,
\label{Eq:tensorv}
\eea
where $V(S)$ is the potential of either the inflaton field or the SM Higgs boson, with $S = \phi, H$, and $A\lrd_\mu B\equiv A\partial_\mu B-(\partial_\mu A)B$. The production of spin-$\frac32$ particles was recently considered in \cite{kkmov}.

For the gravitational processes
\beq
\phi/{\rm{SM}}^i(p_1)+\phi/{\rm{SM}}^i(p_2) \rightarrow A_\mu(p_3)+ A_\mu(p_4) \, ,
\label{Eq:process}
\eeq
we parametrize the scattering amplitudes for the production rate of massive vectors as
\begin{equation}
\mathcal{M}^{i 1} \propto M_{\mu \nu}^1 \Pi^{\mu \nu \rho \sigma} M_{\rho \sigma}^i \;, 
\label{scamp}
\end{equation}
where $i=0,1/2,1$ denote the initial state spins involved in the scattering process. 
Here, $\Pi^{\mu \nu \rho \sigma}$ is the graviton propagator for the canonical field $h_{\mu \nu}$ with momentum $k = p_1+p_2$,
\begin{equation}
 \Pi^{\mu\nu\rho\sigma}(k) = \frac{\eta^{\mu\rho}\eta^{\nu\sigma}
 +\eta^{\mu\sigma}\eta^{\nu\rho} 
 - \eta^{\mu\nu}\eta^{\rho\sigma} }{2k^2} \, .
\end{equation} 
 The partial amplitudes, $M_{\mu \nu}^i$, can be expressed by \cite{cmov}
\bea 
\label{partamp1}
M_{\mu \nu}^{0} &=& \frac{1}{2}\left[p_{1\mu} p_{2\nu} + p_{1\nu} p_{2\mu} - \eta_{\mu \nu}p_1\cdot p_2 - \eta_{\mu \nu} V''(S)\right] \,, \\ 
\label{partamp2}
M_{\mu \nu}^\frac12 &=&  \frac{1}{4} {\bar v}(p_2) \left[ \gamma_\mu (p_1-p_2)_\nu + \gamma_\nu (p_1-p_2)_\mu \right] u(p_1) \, , \\
\label{partamp3}
M_{\mu \nu}^{1} &=& \frac{1}{2} \bigg[ \epsilon_{2}^{*} \cdot \epsilon_{1}\left(p_{1 \mu} p_{2 \nu}+p_{1 \nu} p_{2 \mu}\right)
-\epsilon_{2}^{*} \cdot p_{1}\left(p_{2 \mu} \epsilon_{1 \nu}+\epsilon_{1 \mu} p_{2 \nu}\right) - \epsilon_{1} \cdot p_{2}\left(p_{1 \nu} \epsilon_{2 \mu}^{*}+p_{1 \mu} \epsilon_{2 \nu}^{*}\right)~\nonumber \\
&+& \left(p_{1} \cdot p_{2} + m_{A}^2 \right)\left(\epsilon_{1 \mu} \epsilon_{2 \nu}^{*}+\epsilon_{1 \nu} \epsilon_{2 \mu}^{*}\right) +\eta_{\mu \nu}\left(\epsilon_{2}^{*} \cdot p_{1} \epsilon_{1} \cdot p_{2}-\left( p_{1} \cdot p_{2} + m_{A}^2 \right) \, \epsilon_{2}^{*} \cdot \epsilon_{1}\right) \bigg]  \, .
\eea
Here we keep our discussion completely general and argue that any inflationary model that agrees with the slow-roll parameters determined by \textit{Planck} data~\cite{Planck} can be used, as long as it has a well-defined minimum and the potential can be expanded about its minimum as given by Eq.~(\ref{kexp}). Therefore, the currently favored Starobinsky model~\cite{Staro} (for $k=2$) and $\alpha$-attractor models~\cite{Kallosh:2013hoa} (for arbitrary $k$) are perfectly sufficient~\cite{egnov}.

\subsection{Gravitational production from the inflaton}
\label{procaprod}

The gravitational production of massive vector dark matter processes is illustrated by the Feynman diagram in Fig.~\ref{Fig:feynman}.
We consider first the process of vector dark matter production from the inflaton condensate $\phi + \phi \rightarrow A_\mu + A_\mu$. For the case of a quadratic potential, the inflaton behaves like a massive particle at rest with four-momentum $p_{1,2}$.\footnote{This statement holds true provided the symmetry factors are taken into account. For a detailed discussion, see \cite{MO,cmov}.} We compute the partial amplitude using Eq.~(\ref{partamp1}) with $V(\phi)$ 
given by Eq.~(\ref{kexp}) and use the inflaton condensate zero mode. For more details, see~\cite{cmosv}. The dominant particle production typically occurs right after the end of inflation when the amplitude of the oscillation is maximal. The gravitational particle production is Planck-suppressed, but the vector dark matter production from the inflaton is still significant since the energy density of the inflaton is huge at the beginning of reheating and a significant amount of it is transferred during the beginning of the reheating process.

\subsubsection{Quadratic potential minimum}
To begin, we first focus on the simplest case with a quadratic potential minimum given by 
$V(\phi) \simeq \frac{1}{2} m_{\phi}^2 \phi^2$. To compute the rate, we first find the square of the matrix element in Eq.~(\ref{scamp}) and use $M^0_{\rho\sigma}$
for the incoming state of the inflaton. For the inflaton condensate, we assume that the incoming 3-momentum of the inflaton vanishes, with $\Vec{p}_{1, 2} = 0$, and proceed to compute the amplitude element squared $|\mathcal{M}^{01}|^2$ by summing over all polarizations of the outgoing vector states. We find that the matrix element squared is given by:\footnote{We 
note that we include the symmetry 
factors associated with identical 
initial and final states in the 
squared amplitude, $|\overline{\mathcal{M}}|^2$.}
\begin{equation}
    \label{matelsq1}
|\overline{\mathcal{M}}|^2 \; = \; \frac{3 m_A^4 \left(1 - \frac{2 m_{\phi}^2}{s} \right)^2 - 2m_A^2 m_{\phi}^2 \left(1 - \frac{2 m_{\phi}^2}{s} \right) + m_{\phi}^4}{4M_P^4}\; = \; \frac{1}{16} \frac{m_{\phi}^4}{M_P^4} \left(4 - 4\tau + 3\tau^2 \right)\, ,
\end{equation}
where in the second equality we used the Mandelstam variable $s = 4 m_{\phi}^2$ and we introduced the definition $\tau \equiv m_A^2/m_{\phi}^2$. 
It is worth noting that the massless limit of the result in Eq.~(\ref{matelsq1}) is finite.
While there is no gravitational 
production of massless fermions due to helicity 
conservation, or to massless gauge bosons due to conformal invariance \cite{ema,MO,cmov}, the production of spin-$\frac{3}{2}$ particles (raritrons) diverges in the massless limit \cite{kkmov} due to the pathology of spins greater than one.
As we will see, the massless limit of (\ref{matelsq1}) corresponds to the amplitude for producing a real scalar. We will return to this question when we  consider the Stueckelberg mechanism for massive vectors in Section \ref{stueck}. 

Note that the above expression for $|\overline{\mathcal{M}}|^2$ is summed over
both transverse and longitudinal modes. 
We can
distinguish between the production of transverse ($\pm 1$) and 
longitudinal ($0$) modes by
introducing the following forms of the polarization vectors
\begin{align}
& \epsilon_{+}^\mu(p)=\frac{1}{\sqrt{2}}\left(\begin{array}{c}
0 \\
-\cos \theta \cos \phi+i \sin \phi \\
-\cos \theta \sin \phi-i \cos \phi \\
\sin \theta
\end{array}\right) \, ,\\
& \epsilon_0^\mu(p)=\frac{1}{m_A}\left(\begin{array}{c}
|\boldsymbol{p}| \\
E \sin \theta \cos \phi \\
E \sin \theta \sin \phi \\
E \cos \theta
\end{array}\right) \, , \\
& \epsilon_{-}^\mu(p)=\frac{1}{\sqrt{2}}\left(\begin{array}{c}
0\\\cos \theta \cos \phi+i \sin \phi \\
\cos \theta \sin \phi-i \cos \phi \\
-\sin \theta
\end{array}\right) \, .
\end{align}
The relative contributions to $|\overline{\mathcal{M}}|^2$ break down as: $2 \tau^2$ for the transverse mode and $(2-\tau)^2$ for the longitudinal mode, each with the same prefactor of $(m_\phi/2M_P)^4$.  Thus we see that for massless gauge bosons, the transverse contribution vanishes due to the conformal structure of the gauge kinetic term. The remaining contribution in the massless limit found the longitudinal mode corresponds to the scalar degree of freedom which is produced in the massless limit \cite{Garny:2015sjg,cmov}.  

Using Eq.~(\ref{matelsq1}), we find that the production rate, $R^{\phi^k}$, for a quadratic minimum with $k = 2$, can be expressed as~\cite{mybook}
\beq
R^{\phi^2} \; = \; 
n_\phi^2\langle \sigma v \rangle =
\frac{\rho_\phi^2}{m_\phi^2}\frac{|{\overline{\cal M}}|^2}{32\pi m_\phi^2} \frac{p_3}{m_\phi}  \, ,
\label{rate2}
\eeq
where the outgoing momentum is given by $p_3 = \sqrt{m_\phi^2 - m_A^2}$, and combining it with the matrix element squared~(\ref{matelsq1}), one  obtains \cite{Ahmed:2021fvt}
\beq
R^{\phi^2} \; =  \frac{2 \times \rho_\phi^2}{256 \pi M_P^4}  \left(1 - \tau + \frac{3}{4}\tau^2 \right) \left(1 - \tau\right)^{1/2} \, .
\label{inflrate}
\eeq
Here we included the factor of $2$ in the numerator to explicitly account for the $2$ produced massive vectors per annihilation.\footnote{Note that our result differs from that obtained in \cite{Barman:2021ugy,Barman:2022tzk}.}

\subsubsection{General Potentials}
As noted earlier, the gravitational production of the massive vector bosons from the inflaton condensate depends on the shape of the potential minimum. We discuss the derivation of the Boltzmann equation for a decaying inflaton condensate in Appendix~\ref{ap:boltz}. Therefore, in this subsection, we consider more general potentials of the form (\ref{kexp}). From Eq.~(\ref{Eq:oscillation}), we can express the potential as 
$V(\phi) = V(\phi_0) \cdot \mathcal{P}(t)^k$. When we Fourier expand $\mathcal{P}(t)^k$ as in Eq.~(\ref{PFexpand}), we can write the potential in terms of its corresponding Fourier modes \cite{Ichikawa:2008ne,Kainulainen:2016vzv,GKMO2}
\beq
V(\phi)=V(\phi_0)\sum_{n=-\infty}^{\infty} {\cal P}_{k,n} e^{-in \omega t}
=\langle \rho_\phi \rangle \sum_{n = -\infty}^{\infty} {\cal P}_{k, n} e^{-in \omega t} \, ,
\eeq
where $\langle \rho_\phi \rangle$ is the mean energy density averaged over the oscillations and $\omega$ is the oscillation frequency of $\phi$, given by \cite{GKMO2}
\beq
\label{eq:angfrequency}
\omega \; = \; m_\phi \sqrt{\frac{\pi k}{2(k-1)}}
\frac{\Gamma(\frac{1}{2}+\frac{1}{k})}{\Gamma(\frac{1}{k})} \, ,
\eeq
with $m_\phi^2 = \partial^2 V/\partial \phi^2|_{\phi_0}$. 

To calculate the scattering rate of the inflaton condensate, we follow the approach outlined in Appendix~\ref{ap:boltz}. 
We find that the massive vector production rate is given by
\begin{equation}
    \label{eq:rategenk}
    R^{\phi^k} \; = \; \frac{2 \times \rho_{\phi}^2}{16 \pi M_P^4}  \Sigma_{1}^k \, ,
\end{equation}
with
\begin{align}
    \label{sigma1}
    \Sigma_{1}^k \; = \; \sum_{ n = 1}^{+\infty} |\mathcal{P}_{k, n}|^2 \left(1-4 \frac{m_{A}^2}{E_n^2} + 12 \frac{m_{A}^4}{E_n^4} \right) \times \left[1 - \frac{4m_{A}^2}{E_n^2} \right]^{1/2} \, ,
\end{align}
where $E_n = n \omega$ is the $n$-th mode energy of the inflaton oscillation.
For the $k=2$ case, with $\omega = m_\phi$ (see Eq.~(\ref{eq:angfrequency})) and
${\cal P}(t)^2=\cos^2 (m_\phi t)=\frac{1}{2}+\frac{1}{4}(e^{-2 m_\phi t}+e^{2 m_\phi t})$,
since 
$\sum|\mathcal{P}_{2, n}|^2 =|{\cal P}_{2,2}|^2 = \frac{1}{16}$, only the second mode in the Fourier expansion contributes to the sum.
Using $E_2 = 2 m_{\phi}$, we find that the production rate~(\ref{eq:rategenk}) reduces to Eq.~(\ref{inflrate}).
For other values of $k$, when $m_A \ll m_\phi$, $ \Sigma_{1}^k \simeq \sum |\mathcal{P}_{k, n}|^2$. For example, for $k=4 (6)$, $\Sigma_{1}^k \simeq 0.063~(0.056)$ \cite{cmov}.

\begin{figure}[!ht]
  \centering
\includegraphics[width=0.65\textwidth]{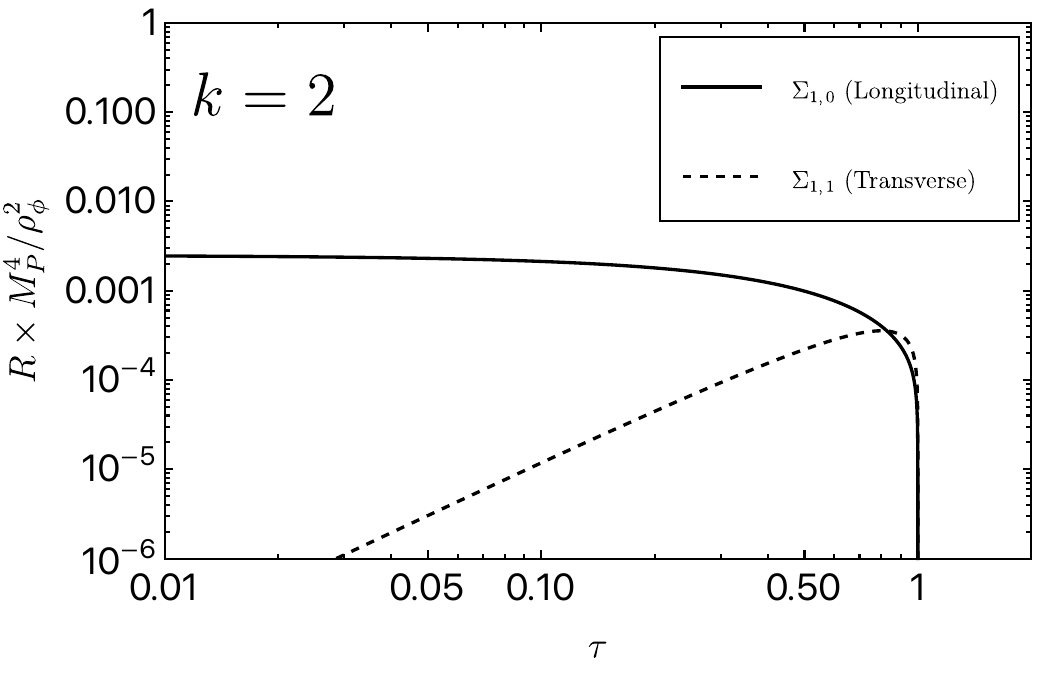}
  \caption{\em \small Longitudinal (solid) and transverse (dashed) gravitational production 
  rates of massive vector bosons for $k=2$ in units of $R \times M_P^4/\rho_{\phi}^2$ as a function of $\tau = m_A^2/m_{\phi}^2$.} 
  \label{Fig:rates}
\end{figure}

The massive vector boson production rate~(\ref{eq:rategenk}) can then be decomposed as
\begin{equation}
    \label{eq:rategenkd}
    R^{\phi^k} \; = \; \frac{2 \times \rho_{\phi}^2}{16 \pi M_P^4}  \left(\Sigma_{1, 1}^k + \Sigma_{1, 0}^k \right)\, ,
\end{equation}
where the transverse spin $\pm 1$ contribution can be expressed as
\begin{align}
    \Sigma_{1, 1}^k \; = \; \sum_{ n = 1}^{+\infty}|\mathcal{P}_{k, n}|^2 \times 8 \frac{m_A^4}{E_n^4} \times \left[1 - \frac{4m_{A}^2}{E_n^2} \right]^{1/2} \, ,
    \label{trans}
\end{align}
and the longitudinal spin $0$ contribution is
\begin{align}
     \Sigma_{1, 0}^k \; = \; \sum_{ n = 1}^{+\infty} |\mathcal{P}_{k, n}|^2 \times\left(1- \frac{2m_{A}^2}{E_n^2} \right)^2 \times \left[1 - \frac{4m_{A}^2}{E_n^2} \right]^{1/2} \, . 
    \label{long}
\end{align}
We note that the sum of transverse and longitudinal components satisfy the expression~(\ref{sigma1}), with $\Sigma_{1}^k = \Sigma_{1, 1}^k + \Sigma_{1, 0}^k$. 
We also find in Eqs.~(\ref{trans}) and (\ref{long}) the 
behaviour we noted above, namely that in the limit $m_A \rightarrow 0$, the production of transverse modes vanishes, and only the longitudinal component is produced gravitationally.
 This result is illustrated in Fig.~\ref{Fig:rates}, which shows the evolution of the transverse and longitudinal production rates as a function of the ratio $\tau$ in the case of the quadratic potential, $k=2$. What is striking is that production of the longitudinal mode 
greatly dominates for $\tau <1$. Already for $\tau = 1/2$, 90\% of the production of the vector is through the longitudinal mode. This agrees with the results found in \cite{Dimopoulos:2006ms,Graham:2015rva,Ahmed:2020fhc,Kolb:2020fwh} using very different techniques. 
Also note that the absolute value of the rates  depends 
on the Fourier coefficients ${\cal P}_{n,k}$, which themselves become very similar for large values 
of $k$, hence we do not expect big differences for larger values of $k$.

\subsection{Relic Abundance}
\label{sec:relic}

To compute the relic abundance of massive vector fields, we start with the Boltzmann equation
\beq
\frac{dn}{dt} + 3Hn = R^{\phi^k} \,,
\label{Eq:boltzmann1}
\eeq
where $n$ is the dark vector number density. If we introduce the yield $Y \equiv a^3 n$, we can rewrite the Boltzmann equation as a function of the scale factor,
\beq
\frac{dY}{da}=\frac{a^2R^{\phi^k}}{H(a)} \, .
\label{Eq:boltzmann2}
\eeq
Assuming that the inflaton energy density dominates the total density, we can write
\beq
H(a)=\frac{\rho^\frac{1}{2}_\phi(a)}{\sqrt{3}M_P} \, ,
\eeq
and $\rho_\phi(a)$ given by Eq.(\ref{Eq:rhophi}), the Boltzmann equation~(\ref{Eq:boltzmann2}) can be expressed as
\beq
\frac{dY}{da}=\frac{\sqrt{3}M_P}{\sqrt{\rho_{\rm RH}}}a^2\left(\frac{a}{a_{\rm RH}}\right)^{\frac{3k}{k+2}}R^{\phi^k}(a) \, .
\label{Eq:boltzmann4}
\eeq
We first focus on the simplest case $k = 2$, which leads to $\rho_\phi \sim a^{-3}$, $\rho_R \sim T^4 \sim a^{-3/2}$, and the inflaton mass $m_\phi^2 = 2 \lambda M_P^2$. Using $R^{\phi^2}(a)$ given by Eq.~(\ref{inflrate}), 
Eq.~(\ref{Eq:boltzmann4}) is easily integrated and we obtain
\begin{eqnarray}
n(a_{\rm RH}) \; = \; \frac{1}{64 \sqrt{3} \pi  M_P} \left(\frac{\rho_{\rm end}}{M_P^4} \right)^\frac12 \frac{g_{\rm RH} \pi^2}{30} T_{\rm RH}^4 \times \left(1 - \tau + \frac{3}{4}\tau^2 \right) \left(1 - \tau\right)^{1/2} \, ,
\label{n32tot}
\end{eqnarray}
where we assumed that $a_{\rm RH} \gg a_{\rm end}$ and $g_{\rm RH}$ is the number degrees of freedom at reheating. We can then compute the relic abundance~\cite{mybook}
\beq
\label{eq:relicabund1}
\Omega h^2 \; \simeq \; 1.6\times 10^8\frac{g_0}{g_{\rm RH}}\frac{n(\trh)}{\trh^3}\frac{m_{A}}{1~{\rm GeV}} \, ,
\eeq
and find
\begin{eqnarray}
\Omega h^2 & \simeq & 0.12 \times \left( \frac{T_{\rm RH}}{10^{10} {\rm GeV}} \right) \left( \frac{\rho_{\rm end}}{(5.2 \times 10^{15} {\rm GeV})^4} \right)^\frac12 \left( \frac{m_A}{10^7 \, \rm GeV} \right) \, ,
\label{oh2tot}
\end{eqnarray}
where we used $g_0=43/11$ and $g_{\rm RH}=915/4$, together with the assumption $m_{A} \ll m_\phi$. Here, we normalized the values of $m_\phi$ and $\rho_{\rm end}$ for an $\alpha$-attractor model of inflation with $k=2$, though it weakly depends on $T_{\rm RH}$, see Refs.~\cite{GKMO1,cmov,egnov} for a detailed discussion.

We can compare the result obtained in (\ref{oh2tot}) 
with that obtained for the relic density of scalars \cite{MO,cmov}, fermions \cite{MO,cmov}, or raritrons \cite{kkmov}. For scalar dark matter, the relic density is approximately the same 
as the expression in Eq.~(\ref{n32tot}) for vectorial dark matter (one need only replace $(1-\tau + \frac34 \tau^2) \to (1+\tau)^2$) because it is the longitudinal mode which 
is mainly produced when $T_{RH}\lesssim m_\phi$. They are identical when in the limit $\tau \to 0$.  In contrast, the fermion ($\chi$) relic density is suppressed by a factor proportional to 
$m_\chi^2/m_\phi^2$ (in this case, one need only replace $(1-\tau + \frac34 \tau^2) \to (\frac12\tau^2)$). Thus 
fermion masses 
$10^4$ times larger are required before saturating the matter content of the Universe.
On the other hand, the relic density of a raritron of mass $10^7$ GeV 
leads to $\Omega_{3/2} h^2 \gtrsim 10^{11}$ \cite{kkmov} 
due to the very efficient production of longitudinal modes with $R^{\phi^2}_{3/2}\propto m_\phi^2/m_{3/2}^2$.

From Eqs. (\ref{eq:rategenkd}) - (\ref{long}) we can repeat the
exercise of separating the production of transverse and longitudinal modes.
 For the transverse contribution for $k=2$, we find
\beq
n_1(a_{\rm RH}) \; = \; \frac{1}{64 \sqrt{3} \pi  M_P} \left(\frac{\rho_{\rm end}}{M_P^4} \right)^\frac12 \alpha T_{\rm RH}^4 \times \left(\frac{1}{2} \tau^4 \right) \left(1 - \tau\right)^{1/2} \, ,
\eeq
and
\begin{eqnarray}
\Omega_1 h^2 & \simeq & 7 \times 10^{-17} \left( \frac{T_{\rm RH}}{10^{10}~ {\rm GeV}} \right) \left( \frac{\rho_{\rm end}}{(5.2 \times 10^{15} {\rm GeV})^4} \right)^\frac12 \left( \frac{1.7 \times 10^{13} {\rm GeV}}{m_\phi} \right)^4 \left( \frac{m_{A}}{{\rm EeV}} \right)^5 \, .
\label{oh2trans}
\end{eqnarray}
As expected and discussed previously, the gravitational production of the transverse mode is completely negligible. Similarly, we can compute the longitudinal contribution
\begin{eqnarray}
n_{0}(a_{\rm RH}) & = & \frac{1}{64 \sqrt{3} \pi  M_P} \left(\frac{\rho_{\rm end}}{M_P^4} \right)^\frac12 \alpha T_{\rm RH}^4 \times \left(1 -\tau^2 \right) \left(1 - \tau\right)^{1/2} \, ,
\end{eqnarray}
which for
$m_{A} \ll m_\phi$, gives the result in Eq.~(\ref{n32tot}) since the production of massive vector field is 
completely dominated by the longitudinal component
and
\begin{eqnarray}
\Omega_0 h^2 & \simeq & 0.12 \times \left( \frac{T_{\rm RH}}{10^{10} {\rm GeV}} \right) \left( \frac{\rho_{\rm end}}{(5.2 \times 10^{15} {\rm GeV})^4} \right)^\frac12 \left( \frac{m_A}{10^7 \, \rm GeV} \right) \, ,
\label{oh2long}
\end{eqnarray}
which coincides with Eq.~(\ref{oh2tot}).

Finally, we can also generalize this result to cases when $k \neq 2$. We find that the number density can be expressed as
\begin{eqnarray}
    n(a_{\rm RH}) \; = \; \frac{(k+2)}{(k-1)}\frac{\rho_{\rm RH}^{3/2}}{16\sqrt{3} \pi M_P^{3}} \left(\frac{\rho_{\rm end}}{\rho_{\rm RH}}\right)^{1-\frac{1}{k}}\Sigma_{1}^k \, ,
\end{eqnarray} 
which is identical to the result for a scalar with the replacement of $\Sigma_1^k \to \Sigma_0^k$ which are in fact equal in the limit $m_A \to 0$ \cite{cmov}.
The dark matter abundance becomes
\begin{align}
&\Omega h^2\simeq 9.7 \times 10^{5}\frac{(k+2)}{(k-1)} \left(\frac{\rho_{\rm end}}{\rho_{\rm RH}} \right)^{1-\frac{1}{k}} \frac{\rho_{\rm RH}^{3/4}}{M_P^3} \frac{m_{A}}{1~{\rm GeV}} \Sigma_{1}^k  \, .
\label{oh2totgenk1}
\end{align}
As expected, for $k=2$ this expression reduces to Eq.~(\ref{oh2tot}).

We illustrate these results in Fig.~\ref{Fig:dmconst} where we show the lines of constant $\Omega_1 h^2$ on the $(m_A, \trh)$ plane.{\footnote{In a supersymmetric theory, one should also consider the production of gauginos. However the production rate for gauginos is suppressed by $m_\lambda^2/m_\phi^2$ and is negligible for $m_A \sim m_\lambda \ll m_\phi$.} The line for $k=2$, corresponds roughly to $m_A \trh \lesssim 10^{17}$~GeV$^2$. For $k=4$, there is no dependence on the reheating temperature, and there is an upper limit to the vector mass, $m_A \lesssim 120$~GeV. For $k=6$, we have $m_A/\trh^{1/3} \lesssim .0017$~GeV$^{2/3}$. We also note that our analysis only captures the post-inflationary production, corresponding to the UV modes in the particle number abundance spectrum. For a more detailed discussion that includes the IR particle power spectrum contribution arising during inflation, see Refs.~\cite{Graham:2015rva,Ema:2019yrd,Kolb:2020fwh}.

\begin{figure}[!ht]
\centering
\includegraphics[width=0.6\textwidth]{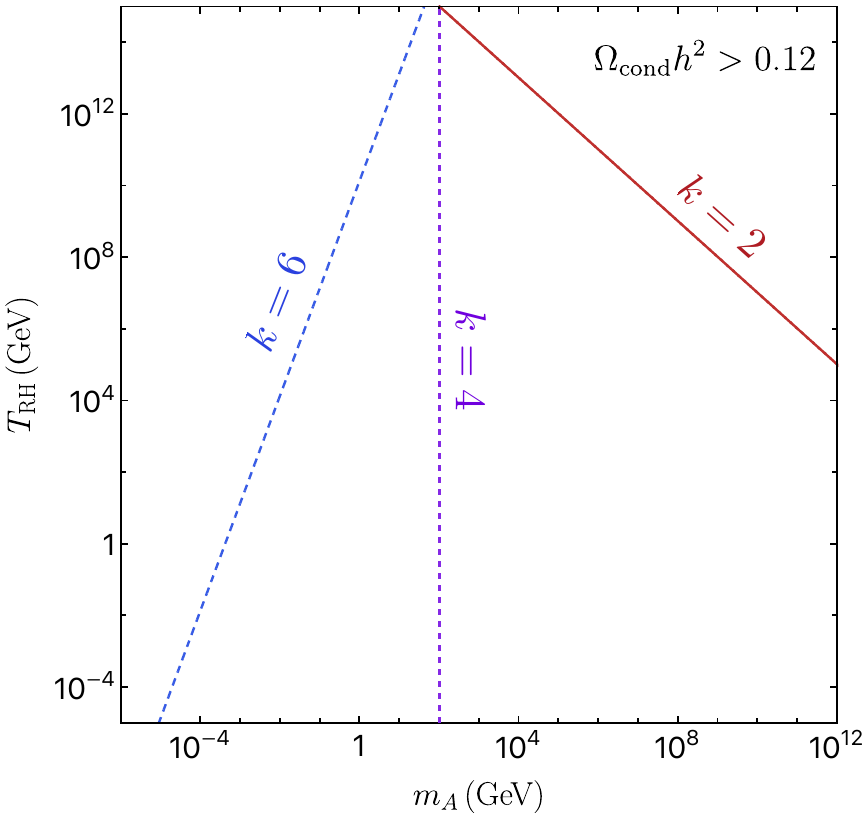}
\caption{\em \small 
Contours of $\Omega h^2 = 0.12$ on the $(m_A,\trh)$ plane for $k=2, 4$, and 6.
}
\label{Fig:dmconst}
\end{figure}

\subsection{Thermal Production of Massive Vector Dark Matter}
\label{sec:relth}

A second production mechanism for producing massive vector dark matter arises from the Standard Model thermal background, $\textrm{SM} + \textrm{SM}  \rightarrow A_\mu+ A_\mu$, also depicted in Fig.~\ref{Fig:feynman}, and uses the partial amplitudes~(\ref{partamp1})-(\ref{partamp3}) for SM particles on the right-hand side of Eq.~(\ref{scamp}). The scattering processes of SM particles include the Higgs scalars, gauge bosons, and fermions in the initial state. Given our assumption of the MSSM, massless superpartners are also included in these initial states. Since the initial particle momenta $p_1$ and $p_2$ are large (of order $m_\phi$) at the beginning of reheating, dominating over electroweak scale quantities, we can assume these initial particle states to be effectively massless. We provide a detailed computation of the thermal rates in Appendix~\ref{ap:ampandthermal}.

For scalar initial states, including Higgses and sfermions, we use Eq.~(\ref{fullM2}) with the association $\phi \to s$ and 
set $m_\phi = 0$ (i.e., we neglect the scalar masses, and all 
MSSM masses relative to the reheating temperature in the 
thermal bath). Similarly, we use Eqs.~(\ref{amp:vect2}) and (\ref{amp:vect3}) for the initial massless fermions and gauge bosons. To calculate the production rate, we integrate the amplitude squared over the incoming states
\cite{Benakli:2017whb,grav2,mybook},
\begin{eqnarray}
R(T) & = & \frac{2}{1024 \pi^6}\times\int f_1 f_2 E_1 \diff E_1 E_2 \diff E_2 \diff \cos \theta_{12} \, |\overline{\cal M}|^2 \diff \Omega_{13} \, ,
\label{thrate}
\end{eqnarray}
where we assume that $s \gg 4m_A^2$,  with $s = (p_1 + p_2)^2$. Here the factor of two accounts for two massive vectors produced per scattering, $E_i$ denotes the energies of the initial and final state particles,
$\theta_{13}$ and $\theta_{12}$ are the angles formed by momenta ${\bf{p}}_{1,3}$ (in the center-of-mass frame) and ${\bf{p}}_{1,2}$ (in the laboratory frame), respectively, and $\diff\Omega_{13}=2\pi \diff\cos\theta_{13}$.  We assume the following thermal distributions for the incoming particles
\beq
f_{i} \; = \; \frac{1}{e^{E_i/T}\pm 1} \, ,
\eeq 
and obtain the expression for $|\overline{\mathcal{M}}|^2$ by summing over the three amplitudes associated with the different spins of the initial states
\begin{equation}
    \label{Eq:ampscat}
    |\overline{\mathcal{M}}|^{2}=  N_b |\overline{\mathcal{M}}^{0}|^{2}+ N_f|\overline{\mathcal{M}}^{1 / 2}|^{2}+N_V|\overline{\mathcal{M}}^{1}|^{2} \, ,
\end{equation}
where $N_b, N_f$, and $N_V$ represent the number of bosonic, fermionic, and vector fields in the model, respectively. For SM initial states, these values are $N_b = 4$, $N_f = 45$, and $N_V = 12$. For MSSM initial states, which we focus on here, the corresponding values are $N_b = 98$, $N_f = 61$, and $N_V = 12$. The matrix elements and their computation is detailed in Appendix \ref{ap:ampandthermal}. The gravitational production rate can then be expressed as 
\beq
R^T_{1} \; = \; \beta_1 \frac{T^{8}}{M_P^4} 
+ \beta_2 \frac{m_{A}^2 T^6}{M_P^4} + \beta_3 \frac{m_{A}^4 T^4}{M_P^4}  \, ,
\label{Eq:ratethermal}
\eeq
with the coefficients $\beta_i$ given by Eqs. (\ref{Eq:beta1}-\ref{Eq:beta3}) for SM and Eqs. (\ref{Eq:mssmbeta1}-\ref{Eq:mssmbeta3}) for MSSM.

Since the thermal production rate is strongly sub-dominant compared to the 
production from the inflaton condensate, we consider only $k=2$. To compute the thermally-produced vector number density, we 
replace the rate in Eq.~(\ref{Eq:boltzmann4}) by the thermal production rate~(\ref{Eq:ratethermal}).
The temperature can be expressed as function of the scale 
factor by solving
\beq
\frac{d \rho_R}{da}+4\frac{\rho_R}{a}=\frac{\Gamma_\phi \rho_\phi}{Ha}\,.
\eeq
We find that the thermally-produced massive vector number density is given by
\begin{align}
n^T(T_{\rm RH}) \; = \; \frac{2 \beta_1}{\sqrt{3}} \frac{\rho_{\rm RH}^{3/2}}{M_P^3} +\frac{4 \beta_2}{3\sqrt{3} \alpha^{3/2}} \frac{\rho_{\rm RH} m_{A}^2}{M_P^3} + \frac{\beta_3 }{\sqrt{3} \alpha} \frac{\sqrt{\rho_{\rm RH}}m_{A}^4}{M_P^3} \,.
\label{Eq:nthermal}
\end{align}    
As before, we assumed that $4m_A^2 \ll s$, which approximately corresponds to $m_A \lesssim T_{\rm RH}$, and we integrated Eq.~(\ref{Eq:boltzmann4}) between $a_{\rm end}$ and $a_{\rm RH}$. 
Since the $\beta_1$ terms dominates the thermal production when $m_A \ll T_{\rm RH}$, we only keep this term, and using Eq.~(\ref{eq:relicabund1}) for the relic abundance, we obtain
\bea
    \Omega^T h^2 & = & 2.7 \times 10^6 \, \frac{n^T(T_{\rm RH})}{T_{\rm RH}^3} \frac{m_{A}}{1~{\rm GeV}} \; \simeq \; 1.7 \times 10^{-10} \left( \frac{T_{\rm RH}}{10^{10}~ {\rm GeV}} \right)^3 \left( \frac{m_A}{10^7{\rm GeV}} \right) \, .
    \label{Oh2therm}
\eea
Comparing Eq.~(\ref{Oh2therm}) with Eq.~(\ref{oh2tot}), we conclude that thermal production is negligible compared to the production from the inflaton condensate for $k=2$. The result is also valid for larger values of $k$. Consequently, the relic density generated by the thermal source is negligible compared with that generated by inflaton oscillations across the entire parameter space. This result remains valid for $k >2$.

\subsection{The Vector Mass from the Stueckelberg and Higgs Mechanisms}
\label{stueck}

Our discussion so far focused on a massive vector field described by the Proca Lagrangian \eqref{procaL}, with a generic mass term. Despite its simplicity, the Proca Lagrangian has several disadvantages.   Contrary to the massless case, the gauge invariance is explicitly broken by the mass term and one can  not smoothly apply the massless limit as one loses one degree of freedom in the massless limit. On the other hand, the mismatch of the degrees of freedom can also be seen when we consider the squared amplitude for the production of  $A_\mu$, e.g., Eq.~\eqref{matelsq1}, when the vector mass is sent to 0.
An alternative description of a massive vector field is  given by Stueckelberg \cite{Stueckelberg:1938hvi,Ruegg:2003ps}. In fact, the Proca Lagrangian can be made gauge invariant through  the field redefinition:
\begin{equation}
    A_\mu \rightarrow A_\mu -\frac{1}{m_A}\partial_\mu\sigma \, ,
\end{equation}
so we get
\begin{equation}
\begin{aligned}
     \mathcal{L} \; = \; -\frac{1}{4}F^{\mu\nu}F_{\mu\nu}+\frac{1}{2}m_A^2 A^\mu A_\mu+\frac{1}{2}\partial_\mu\sigma\partial^\mu\sigma-m_AA^\mu {\partial_\mu}\sigma \, .
\end{aligned}\label{stuckL}
\end{equation}
The new Lagrangian is gauge invariant under the transformations
\begin{equation}
A_\mu\rightarrow A_\mu +\partial_\mu\Lambda,\quad  \sigma\rightarrow \sigma+m_A\Lambda \, .
\label{gaugeinv-stuck}
\end{equation}
We easily recover the Proca Lagrangian by setting the unitary gauge $\sigma=0$. On the other hand, the massless limit now gives rise to a massless vector field along with a scalar, hence the number of degrees of freedom is conserved. Indeed, the new scalar corresponds to the longitudinal mode of the massive vector, that drops out when $m_A\rightarrow0$.

To quantize the theory, we add to the Stueckelberg Lagrangian \eqref{stuckL} a gauge-fixing term:
\begin{equation}
    \mathcal{L}_\text{gf} \; = \; -\frac{1}{2\xi}\left(\partial^\mu A_\mu +\xi m_A\sigma\right)^2 \, ,
\end{equation}
from which we recover a massive vector field and a decoupled massive scalar. The  Feynman gauge choice amounts to $\xi=1$ and the unitary gauge corresponds to $\xi\rightarrow \infty$. We still have three degrees of freedom because, in addition to the subsidiary condition $\partial^\mu A_\mu=-\xi m_A \sigma$, the  gauge invariance \eqref{gaugeinv-stuck} also holds, but now the gauge parameter must satisfy $\left(\partial^2 +\xi m_A^2\right)\Lambda=0$. The Stueckelberg mechanism can allow for, e.g., a massive  abelian gauge boson whose mass is not provided by a Higgs boson through spontaneous symmetry breaking. Examples of the Stueckelberg extension to the SM or supersymmetry can be found in \cite{Kors:2004dx,Kors:2004iz,Kors:2005uz}.

Assuming minimal coupling to gravity,  it is possible to  generalize this Lagrangian to the curved space \cite{Belokogne:2015etf} and linearly perturb the metric around Minkowski, as we have done in the beginning of this section, then the  canonically-normalized graviton field $h_{\mu\nu }$  will not only couple to $A_\mu$, but also to $\sigma$. For simplicity, we choose the Feynman gauge $\xi=1$, so the action becomes
\begin{eqnarray}
    S \; = \; \int d^4x \sqrt{-g}\left(-\frac{1}{4}F^{\mu\nu}F_{\mu\nu}+\frac{1}{2}m_A^2 A^\alpha A_\alpha-\frac{1}{2}(\nabla^\alpha A_\alpha)^2+\frac{1}{2}\nabla^\mu\sigma\nabla_\mu\sigma-\frac{1}{2}m_A^2 \sigma^2\right) \, .
    \label{stuck-feynman}
\end{eqnarray}
The Stueckelberg scalar is decoupled and has the same mass as the vector boson. Note also that in this gauge, the polarization sum for $A_\mu $ reads $\sum_\lambda \epsilon_{\lambda,\mu}\epsilon_{\lambda,\nu}^*=-g_{\mu\nu}$, as if $A_\mu$ was ``massless''.
In the Minkowski limit, the energy-momentum tensor is:
\begin{equation}
    T^{\mu\nu} \; = \; T_\sigma^{\mu\nu}+T_A^{\mu\nu} \, ,
\end{equation}with\begin{equation}
    \begin{aligned}
        T_\sigma^{\mu\nu} \; = \; \partial^\mu\sigma\partial^\nu\sigma-\eta^{\mu\nu} \left(\frac{1}{2}\partial^\alpha\sigma\partial_\alpha\sigma-\frac{1}{2}m_A^2\sigma^2 \right) \, ,
    \end{aligned}
\end{equation}\begin{equation}
    \begin{aligned}
        T_A^{\mu\nu}=&-\frac{1}{2}F^{\mu\alpha}F^\nu{}_\alpha-\frac{1}{2}F^{\nu\alpha}F^\mu{}_\alpha+m_A^2A^\mu A^\nu+2A^{(\mu} \partial^{\nu )}\partial^\alpha A_\alpha\\&-\eta^{\mu\nu} \left(-\frac{1}{4}F^{\alpha\beta}F_{\alpha\beta}+\frac{1}{2}m_A^2A^\alpha A_\alpha+\frac{1}{2}(\partial^\alpha A_\alpha)^2+A^\beta\partial_\beta \partial_\alpha A^\alpha \right) \, .
    \end{aligned}
\end{equation}
One can easily check the conservation  $\partial_\nu T^{\mu\nu}=0$ upon using the Klein-Gordon equations. The energy-momentum tensor for $\sigma$ is well-known and is identical to that in Eq.~(\ref{Eq:tensors}) when $V(\sigma) = \frac12 m_A^2 \sigma^2$. In contrast, the energy-momentum tensor for $A_\mu $ is different from that derived from the Proca Lagrangian, Eq.~\eqref{Eq:tensorv}. In the following, we compute the production rate of  Stueckelberg vector field $A_\mu$ from the process in Fig.~\ref{Fig:feynman}, and compare that to the production of the Proca vector field $A_\mu$.

The partial amplitude for $A_\mu$ is now given by
\begin{equation}
    \begin{aligned}
        M_{\mu \nu}^{1} =& \frac{1}{2} \bigg[ \epsilon_{2}^{*} \cdot \epsilon_{1}\left(p_{1 \mu} p_{2 \nu}+p_{1 \nu} p_{2 \mu}\right)
-\epsilon_{2}^{*} \cdot p_{1}\left(p_{2 \mu} \epsilon_{1 \nu}+\epsilon_{1 \mu} p_{2 \nu}\right) - \epsilon_{1} \cdot p_{2}\left(p_{1 \nu} \epsilon_{2 \mu}^{*}+p_{1 \mu} \epsilon_{2 \nu}^{*}\right)~\nonumber \\
&+ \left(p_{1} \cdot p_{2} + m_{A}^2 \right)\left(\epsilon_{1 \mu} \epsilon_{2 \nu}^{*}+\epsilon_{1 \nu} \epsilon_{2 \mu}^{*}\right) +\eta_{\mu \nu}\left(\epsilon_{2}^{*} \cdot p_{1} \epsilon_{1} \cdot p_{2}-\left( p_{1} \cdot p_{2} + m_{A}^2 \right) \, \epsilon_{2}^{*} \cdot \epsilon_{1}\right) \bigg]  \\&+\frac{1}{2}\eta^{\mu\nu}\left[\left(p_1\cdot\epsilon_1\right)\left(p_2\cdot\epsilon_2^*\right)+\left(p_2\cdot\epsilon_1\right)\left(p_2\cdot\epsilon_2^*\right)+\left(p_1\cdot\epsilon_2\right)\left(p_1\cdot\epsilon_1^*\right)\right]\\&-\frac{1}{2}\left[\epsilon_1^\mu p_2^\nu \left(p_2\cdot\epsilon^*_2\right)+\epsilon_2^\mu p_1^\nu \left(p_1\cdot\epsilon^*_1\right)+\epsilon_2^\nu p_1^\mu\left(p_1\cdot\epsilon_1^*\right)\right] \, .
    \end{aligned}
\end{equation}
After some involved algebra, one can verify the simple relation between squared amplitudes:
\begin{equation}
  \left|\mathcal{M}\right|^2_{\text{Stueckelberg, }A}-\left|\mathcal{M}\right|^2_{\sigma}=\left|\mathcal{M}\right|^2_{\text{Proca, }A} \, ,
\end{equation}
which holds for any massive or massless initial states of spin 0, 1/2, 1. The above relation indicates that the production of the Stueckelberg $A_\mu$ differs from  the Proca $A_\mu$ by a scalar contribution. To understand this point, recall that the Stueckelberg Lagrangian  is obtained from the Proca Lagrangian by a  redefinition of $A_\mu$ and adding a gauge fixing term $-\frac{1}{2}\left(\partial^\mu A_\mu+ m_A \sigma \right)^2$. The  two Lagrangians are equivalent if one requires for the former that the physical states be those for which (see \cite{Ruegg:2003ps}): \begin{equation}
   \left( \partial^\mu A_\mu +m_A\sigma\right) \left|\text{phys}\right> \; = \;  0 \, .
\end{equation}
Therefore, removing from the amplitude $\left|\mathcal{M}\right|^2_{\text{Stueckelberg, }A}$ the unphysical scalar state, which is of mass $m_A$, we obtain $\left|\mathcal{M}\right|^2_{\text{Proca, }A}  $.  Once the unphysical state is removed, the production rate for the vector with a Stueckelberg induced mass is the same as that discussed in Section \ref{procaprod} for the Proca Lagrangian.

Of course, another way to generate a gauge boson mass is through the Higgs mechanism. The relevant Lagrangian is
\begin{equation}
    \mathcal{L}_H \; = \; -\frac{1}{4}F^{\mu\nu}F_{\mu\nu}+|\partial_\mu H-ie A_\mu H |^2-\lambda \left(H^\dagger H -\frac{v^2}{2}\right)^2 \, ,
\end{equation}
where $H$ denotes the complex Higgs scalar. In this case, $A_\mu$ acquires a mass $m_A=e v $ when the Higgs is set to its vacuum expectation value $v/\sqrt{2}$. Going to the unitary gauge and ignoring higher point interactions, one has a gravitational pair production channel for $A_\mu$ and another for $h$ (the physical Higgs). In this case, the squared amplitude for $A_\mu$ production is just $\left|\mathcal{M}\right|^2_{\text{Proca, }A}$. Interestingly, the sum of $A_\mu$ and $h$ productions will coincide with  $ \left|\mathcal{M}\right|^2_{\text{Stueckelberg, }A}$ only if the vector and scalar fields have the same mass. 
Note also that, when $h$ is heavier than $A_\mu$, a secondary production of $A_\mu$ may occur by the decay  or annihilation of the produced $h$, which is absent in the Stueckelberg case.

\section{Summary}
\label{sec:summ}

Inflation is defined by the accelerated expansion of the Universe. To some extent, models of inflation can be phenomenologically triaged using the slow-roll parameters, which are determined during the final stages of the exponential expansion. A model of inflation embedded in a UV theory, which includes the SM (or the MSSM), can be further triaged by its ability to successfully reheat and produce an early stage of radiation domination, and perhaps participate in the process of dark matter production. When examined in detail, the reheating process is not instantaneous and depends on the couplings of the inflaton to matter. Reheating through inflaton decays is most efficient. However the evolution of the radiation bath may depend on the spin of the final state decay products \cite{GKMO2}.

Here we have concentrated on the reheating properties of so-called T-models of inflation \cite{Kallosh:2013hoa}, which yield predictions for the inflationary observables similar to that predicted by the Starobinsky model and are almost independent of $k$ \cite{GKMO2}. For $k=2$
the radiation density scales as $a^{-3/2}$, independent of the spin of the decay product. Reheating to gauge bosons then is relatively efficient \cite{egno4}. However for larger $k$, there can be significant differences, as is easily seen in Table \ref{Tab:table}.
For $k=4$, reheating through the decay to gauge bosons does not occur, though it does for decays to scalars and fermions. 
Note that decays to gauginos also do not allow for reheating for $k=4$. For $k>6$, decays to gauge bosons can lead to reheating, though its efficiency is poor and the reheating temperature is low unless $k$ is relatively large as we have shown in Fig.~\ref{Fig:trehvsk}. 

We have also considered the mechanism of the gravitational production of vectors from the inflaton condensate as well as the thermal bath. The latter was found to be highly sub-dominant. The production of the vectors from the condensate is saturated by the production of the longitudinal mode and is equivalent to the production of a scalar in the limit the vector mass vanishes. We have also discussed the distinction in the production process depending on whether the vector mass is generated by either the Stueckelberg or Higgs mechanism.

\vspace{0.5cm} 
\noindent
\section*{Acknowledgements}
 This project has received support from the European Union's Horizon 2020 research and innovation programme under the Marie Sklodowska-Curie grant agreement No 860881-HIDDeN. M.A.G.G. is supported by the DGAPA-PAPIIT grant IA103123 at UNAM, and the CONAHCYT ``Ciencia de Frontera'' grant CF-2023-I-17. The work of K.K. was supported in part by JSPS KAKENHI No. 20H00160. The work of K.A.O.~was supported in part by DOE grant DE-SC0011842 at the University of Minnesota. The work of S.V. was supported in part by DOE grant DE-SC0022148.
The authors acknowledge the support of the Institut Pascal at Université Paris-Saclay during the Paris-Saclay Astroparticle Symposium 2023, with the support of the P2IO Laboratory of Excellence (program ``Investissements d'avenir'' ANR-11-IDEX-0003-01 Paris-Saclay and ANR-10-LABX-0038), the P2I axis of the Graduate School of Physics of Université Paris-Saclay, as well as IJCLab, CEA, IAS, OSUPS, and the IN2P3 master project UCMN.

\appendix

\section{Effective couplings}
\label{ap:eff}

In this appendix, we describe the derivation of the decay rates arising from the interactions in the Lagrangian~(\ref{gaugekincouplings}) using the approach described in Appendix~\ref{ap:boltz}. In our work, the vector boson is assumed to be massive, though its mass cannot be generated by derivative couplings. We first consider the coupling
\begin{equation}
\mathcal{L} \; \supset \; -\frac{g}{4M_P}\phi F_{\mu\nu}F^{\mu\nu} \, .
\end{equation}
In this case, the amplitude $\mathcal{M}_n$ can be expressed as
\begin{equation}
    \mathcal{M}_n \; = \; -\frac{ig}{2M_P} \phi_0 \mathcal{P}_n \left(p_{A \mu} \epsilon_{A\nu} - p_{A \nu} \epsilon_{A \mu} \right) \left(p^{B \mu} \epsilon^{B \nu} - p^{B \nu} \epsilon^{B \mu} \right) \, ,
\end{equation}
and using the general expression~(\ref{eq:decgeneral}), we obtain
\begin{equation}
  \begin{aligned}  
  \Gamma_{\phi\rightarrow A_\mu A_\mu} &= \frac{g^2 \phi_0^2}{32\pi(1+w_\phi)\rho_\phi M_P^2}\sum_{n=1}^\infty\left|\mathcal{P}_n\right|^2 n \omega \left[6m_A^4 \left<\beta_n(m_A,m_A)_+\right>+n^4\omega^4 \left<\beta_n^3(m_A,m_A)_+\right>\right] \\
  &= \frac{g^2m_\phi^3}{64\pi M_P^2}\frac{(k+2)(k-1)}{m_\phi^5}\sum_{n=1}^\infty\left|\mathcal{P}_n\right|^2 n \omega \left[6m_A^4 \left<\beta_n(m_A,m_A)_+\right>+n^4\omega^4 \left<\beta_n^3(m_A,m_A)_+\right>\right] 
  \\&=\frac{g_{\rm eff}^2m_\phi^3}{64 \pi M_P^2} \, ,
  \end{aligned}
\end{equation}
where we used Eqs.~(\ref{eq:eominf}-\ref{eq:infenden}) and $s = E_n^2 = n^2 \omega^2$. Here we introduced the notation
\begin{equation}
   \begin{aligned}
   &    (x)_+\equiv x\theta(x) \, ,
\\& \beta_n(m_1,m_2)\equiv\sqrt{\left(1-\frac{(m_1+m_2)^2}{E_n^2}\right)\left(1-\frac{(m_1-m_2)^2}{E_n^2}\right)} \, ,
   \end{aligned} 
   \end{equation}
and the effective coupling is given by
\begin{equation}
    g_{\rm eff}^2 \; \equiv \; g^2 \sum_{n=1}^\infty\left|\mathcal{P}_n\right|^2 n \omega \left[6m_A^4 \left<\beta_n(m_A,m_A)_+\right>+n^4\omega^4 \left<\beta_n^3(m_A,m_A)_+\right>\right] \, .
    \label{Eq:geff}
\end{equation}

For $k=2$, we find $\omega = m_{\phi}$ and $\mathcal{P}_1 = 1/2$ (while the components $\mathcal{P}_{n>1} = 0$), and assuming that $m_A \ll m_{\phi}$, we recover $\Gamma_{\phi \rightarrow A_{\mu} A_{\mu}} = \frac{g^2 m_{\phi}^3}{64 \pi M_P^2}$.

Next, we consider the coupling
\begin{equation}
    \mathcal{L} \supset -\frac{\Tilde{g}}{4M_P}\phi F_{\mu\nu}\Tilde{F}^{\mu\nu} \, .
\end{equation}
For this coupling, the transition amplitude $\mathcal{M}_n$ is given by
\begin{equation}
    \mathcal{M}_n \; = \; -\frac{1}{4} \frac{i \tilde{g}}{M_P} \phi_0 \mathcal{P}_n \left(p_{A \mu} \epsilon_{A\nu} - p_{A \nu} \epsilon_{A \mu} \right) \epsilon_{\mu \nu \alpha \beta}\left(p^{B \alpha} \epsilon^{B \beta} - p^{B \beta} \epsilon^{B \alpha} \right) \, ,
\end{equation}
and the decay rate is given by
\begin{equation}
  \begin{aligned}  \Gamma_{\phi\rightarrow A_\mu A_\mu}= &
\frac{\Tilde{g}^2\phi_0^2}{32\pi (1+w_\phi )\rho_\phi M_P^2} \sum_{n=1}^{\infty}
  |\mathcal{P}_n|^2\left(n\omega\right)^5\left<\beta_n^3(m_A,m_A)_+\right>
  \\=&
  \frac{\Tilde{g}^2}{64\pi m_\phi^2M_P^2}(k+2)(k-1) \sum_{n=1}^{\infty}
|\mathcal{P}_n|^2\left(n\omega\right)^5\left<\beta_n^3(m_A,m_A)_+\right> \, , \\
\equiv & \frac{\tilde{g}_{\rm eff}^2 m_{\phi}^3}{64 \pi M_P^2} \, ,
  \end{aligned}
  \end{equation}
where the effective coupling can be expressed as
  \begin{equation}
      \Tilde{g}^2_{{\rm eff}}\equiv \Tilde{g}^2\frac{(k+2)(k-1)}{m_\phi^5} \sum_{n=1}^{\infty}
|\mathcal{P}_n|^2\left(n\omega\right)^5\left<\beta_n^3(m_A,m_A)_+\right> \, .
  \label{Eq:gtildeeff}
  \end{equation}
  When $k=2$ and $m_A \ll m_{\phi}$, we find $\mathcal{P}_1 = 1/2$ and recover $\Gamma_{\phi\rightarrow A_\mu A_\mu}=\frac{\Tilde{g}^2m_\phi^3}{64 \pi M_P^2}$. 

For the coupling
  \begin{equation}
\mathcal{L} \; \supset \; -\frac{\kappa}{4M_P^2}\phi^2 F_{\mu\nu}F^{\mu\nu} \, ,
\end{equation}
we find that the transition amplitude is
\begin{equation}
    \mathcal{M}_{n,m} \; = \; -\frac{i\kappa}{2M_P^2} \phi_0^2 \mathcal{P}_n \mathcal{P}_m \left(p_{A \mu} \epsilon_{A\nu} - p_{A \nu} \epsilon_{A \mu} \right) \left(p^{B \mu} \epsilon^{B \nu} - p^{B \nu} \epsilon^{B \mu} \right) \, ,
\end{equation}
and the scattering rate becomes
\begin{equation}
\begin{aligned}
    \Gamma_{\phi\phi \rightarrow A_\mu A_\mu}=&
\frac{\kappa^2\phi_0^4}{32\pi (1+w_\phi  )\rho_\phi M_P^4 }\sum_{n+m\geq1}|\mathcal{P}_n\mathcal{P}_m|^2 s^{5/2} \left(1 - \frac{4m_A^2}{s} + 6 \frac{m_A^4}{s^2} \right)\sqrt{\left(1-{\frac{4m_A^2}{s}}\right)_+}\\=&
\frac{\kappa^2\phi_0^2(k+2)(k-1)}{64\pi m_\phi^2 M_P^4 }\sum_{n+m\geq1}|\mathcal{P}_n\mathcal{P}_m|^2 s^{5/2} \left(1 - \frac{4m_A^2}{s} + 6 \frac{m_A^4}{s^2} \right)\sqrt{\left(1-{\frac{4m_A^2}{s}}\right)_+} \\=&
\frac{\kappa_{\rm eff}^2 \rho_{\phi} m_{\phi}}{4\pi M_P^4}\, ,
\end{aligned}    
\end{equation}
where $s=(n+m)^2\omega^2$. The effective coupling is given by:
  \begin{equation}
      {\kappa}^2_{{\rm eff}}\equiv{\kappa}^2\frac{k (k+2)(k-1)^2}{16m_\phi^5}\sum_{n+m\geq1}|\mathcal{P}_n\mathcal{P}_m|^2 s^{5/2} \left(1 - \frac{4m_A^2}{s} + 6 \frac{m_A^4}{s^2} \right)\sqrt{\left(1-{\frac{4m_A^2}{s}}\right)_+} \, .
      \label{Eq:kappaeff}
  \end{equation}
When $k=2$ and $m_A \ll m_{\phi}$, we find $\mathcal{P}_1=1/2$ and $|\mathcal{P}_1 \mathcal{P}_1|^2 = 1/16$, and recover $\Gamma_{\phi\phi \rightarrow A_\mu A_\mu}=\frac{\kappa^2\rho_\phi m_\phi}{4\pi M_P^4}$.  

Finally, we consider the coupling
\begin{equation}
\mathcal{L} \; \supset \; -\frac{\Tilde{\kappa}}{4M_P^2}\phi^2 F_{\mu\nu}\Tilde{F}^{\mu\nu} \, ,
\end{equation}
and obtain the transition amplitude
\begin{equation}
    \mathcal{M}_{n, m} \; = \; -\frac{1}{4} \frac{i \tilde{\kappa}}{M_P^2} \phi_0^2 \mathcal{P}_n \mathcal{P}_m \left(p_{A \mu} \epsilon_{A\nu} - p_{A \nu} \epsilon_{A \mu} \right) \epsilon_{\mu \nu \alpha \beta}\left(p^{B \alpha} \epsilon^{B \beta} - p^{B \beta} \epsilon^{B \alpha} \right) \, .
\end{equation}
In this case, the decay rate is given by
\begin{equation}
\begin{aligned}
    \Gamma_{\phi\phi \rightarrow A_\mu A_\mu}=&
\frac{\Tilde{\kappa}^2\phi_0^4}{32\pi (1+w_\phi  )\rho_\phi M_P^4 }\sum_{n+m\geq1}|\mathcal{P}_n\mathcal{P}_m|^2 s^{5/2}{\left(1-\frac{4m_A^2}{s}\right)_+}^\frac{3}{2}\\=&
\frac{\Tilde{\kappa}^2\phi_0^2(k+2)(k-1)}{64\pi m_\phi^2  M_P^4 }\sum_{n+m\geq1}|\mathcal{P}_n\mathcal{P}_m|^2 s^{5/2}{\left(1-\frac{4m_A^2}{s}\right)_+}^\frac{3}{2} \, ,
\end{aligned}    
\end{equation}
where the effective coupling can be expressed as
  \begin{equation}
      \tilde{\kappa}^2_{{\rm eff}}\equiv\tilde{\kappa}^2 \frac{k(k+2)(k-1)^2}{64 m_\phi^5}\sum_{n+m\geq1}|\mathcal{P}_n\mathcal{P}_m|^2 s^{5/2}{\left(1-\frac{4m_A^2}{s}\right)_+}^\frac{3}{2} \, .
      \label{Eq:kappatildeeff}
  \end{equation}
For $k=2$ and $m_A \ll m_{\phi}$, we find $\mathcal{P}_1=1/2$ and $|\mathcal{P}_1 \mathcal{P}_1|^2 = 1/16$, and obtain $\Gamma_{\phi\phi \rightarrow A_\mu A_\mu}=\frac{\Tilde{\kappa} ^2\rho_\phi m_\phi}{4\pi M_P^4}$.

\section{Boltzmann Equation for a Decaying Inflaton Condensate}
\label{ap:boltz}
In this appendix, we describe the derivation of the evolution equation for the energy density of a decaying inflaton condensate. Assuming the inflaton decay is a perturbative
process, the inflaton condensate, $\phi$, is spatially homogeneous, and one can express its phase space distribution (PSD) as $f_{\phi}(k,t) = (2\pi)^3 n_{\phi}(t) \delta^{(3)}(\bf{k})$, where $n_{\phi}$ represents the instantaneous inflaton number density. When neglecting the effects of Bose enhancement or Pauli blocking for the 
decay products of $\phi$, the integrated Boltzmann equation for the number density can be expressed as follows~\cite{Nurmi:2015ema}:
\begin{equation}
\label{eq:boltzm1}
\dot{n}_\phi+3 H n_\phi=-\int d \Psi_{\phi, A, B}|\mathcal{M}|_{\phi \rightarrow A B}^2 f_\phi(k, t) \, .
\end{equation}
Here $A, B$ denote the inflaton decay products, $d \Psi_{\phi, A, B}$ represents the phase space measure for the products $A, B$ and the condensate $\phi$, and $\mathcal{M}$ denotes the transition amplitude. We note that there are no backreaction effects that would produce the inflaton particles in the condensate.
We can express the right-hand side of Eq.~(\ref{eq:boltzm1}) as
\begin{equation}
d \Psi_{\phi, A, B}|\mathcal{M}|_{\phi \rightarrow A B}^2=\sum_{n=1}^{\infty} \frac{d^3 \boldsymbol{k}}{(2 \pi)^3 n_\phi(t)} \frac{d^3 \boldsymbol{p}_A}{(2 \pi)^3 2 p_A^0} \frac{d^3 \boldsymbol{p}_B}{(2 \pi)^3 2 p_B^0}(2 \pi)^4 \delta^{(4)}\left(p_n-p_A-p_B\right)\left|\mathcal{M}_n\right|^2 \, ,
\end{equation}
where $\mathcal{M}_n$ represents the transition amplitude 
for each oscillating field mode of $\phi$ during one oscillation, transitioning from the coherent state $\ket{\phi}$ to a two-particle final state $\ket{A, B}$.
Next, we introduce the following inflaton condensate normalization
\beq
\int \frac{d^3 \boldsymbol{k}}{(2 \pi)^3 n_\phi} f_\phi(k, t) \; = \; 1 \, .
\eeq
Using the above expression and integrating the right-hand side of Eq.~(\ref{eq:boltzm1}), we find
\begin{equation}
\dot{n}_\phi+3 H n_\phi=-\sum_{n=1}^{\infty} \int \frac{d^3 \boldsymbol{p}_A}{(2 \pi)^3 2 p_A^0} \frac{d^3 \boldsymbol{p}_B}{(2 \pi)^3 2 p_B^0}(2 \pi)^4 \delta^{(4)}\left(p_n-p_A-p_B\right)\left|\mathcal{M}_n\right|^2,
\end{equation}
where $p_n = (E_n, \bf{0})$ and $E_n$ represents the $n$-th oscillation mode energy.

The production rate can now be readily obtained from the energy transfer rate from the inflationary to the decay product sector. We introduce the equation of state parameter $w_\phi = p_{\phi}/\rho_{\phi}$ for the inflaton field, and the energy density evolution of $\rho_\phi$ can be expressed as
\beq
\frac{d\rho_{\phi}}{dt} + 3H(1+w_{\phi})\rho_{\phi} \;=\; -(1+w_{\phi})\Gamma_{\phi}\rho_{\phi}\,.
\eeq

Here the right-hand side can be expressed in terms of the energy transfer per space-time volume (Vol$_4$)
\bea
(1+w_\phi)\Gamma_\phi\rho_\phi \; \equiv \; \frac{\Delta E}{{\rm Vol}_4} \,,
\eea
where
\bea
\Delta E 
\; \equiv \; \int\frac{d^3p_A}{(2\pi)^32p_A^0} \frac{d^3p_B}{(2\pi)^32p_B^0}(p_A^0+p_B^0)
\left|\frac{1}{n!}\biggl< {\rm f}\left|\left(i\int d^4x_1{\cal L}_{\rm int}\right)\cdots \left(i\int d^4x_n{\cal L}_{\rm int}\right)\right|0\biggr>\right|^2 \, ,
\label{eq:DeltaE}
\eea
and ${\cal L}_{\rm int}$ is the interaction Lagrangian. The energy transfer rate can be expressed as
\begin{align}
\frac{\Delta E}{{\rm Vol}_4} =
\int\frac{d^3p_A}{(2\pi)^32p_A^0} \frac{d^3p_B}{(2\pi)^32p_B^0}(p_A^0+p_B^0)\sum_{m_1+ \ldots m_n >0}^{\infty}|{\cal M}_{m_1, \ldots ,m_n}|^2(2\pi)^4\delta^4 \left(\sum_i p_{\phi, m_i}-p_A-p_B \right) \, ,
\end{align}
and this leads to the decay (scattering) rate~\cite{Kainulainen:2016vzv, Ichikawa:2008ne, GKMO2}
\begin{equation}
\label{eq:decgeneral}
\Gamma_\phi=\frac{1}{8 \pi\left(1+w_\phi\right) \rho_\phi}\frac{1}{\mathcal{S}!} \sum_{m_1+ \ldots m_n >0}^{\infty}|{\cal M}_{m_1, \ldots ,m_n}|^2 \times \left( E_{m_1}+ \ldots \, +E_{m_n} \right)\times \beta_{m_1, \ldots ,m_n}(m_A, m_B) \, ,
\end{equation}
\begin{equation}
\beta_{m_1, \ldots ,m_n}(m_A, m_B) \equiv \sqrt{\left(1-\frac{\left(m_A+m_B\right)^2}{(\sum_i E_{m_i})^2}\right)\left(1-\frac{\left(m_A-m_B\right)^2}{(\sum_i E_{m_i})^2}\right)} \, ,
\end{equation}
where $\mathcal{S}$ is a symmetry factor for the final states.

We express the oscillating inflaton field as
\begin{equation}
    \phi(t) \; \simeq \; \phi_0(t) \cdot \mathcal{P}(t) \, ,
    \label{Eq:oscillation}
\end{equation}
where ${\cal P}$ is the oscillatory contribution and $\phi_0$ is a slowly-evolving envelope that redshifts with time. One can approximately treat $\phi_0$ as a constant, given its relatively slow evolution over a single oscillation. Next, we decompose the oscillatory contribution as
\bea
{\cal P}(t) \; = \; \sum_{n=-\infty}^{\infty}{\cal P}_n e^{-in\omega t} \, ,
\label{PFexpand}
\eea
where $\omega$ is the inflaton oscillation frequency. 
For an oscillating inflaton field, one can show that the equation of state parameter is given by~\cite{GKMO2}
\begin{equation}
    \label{eq:eominf}
    w_{\phi} \; = \; \frac{k-2}{k+2} \, .
\end{equation}
The effective inflaton mass during the oscillations about the minimum are given by
\begin{equation}
\label{eq:infeffmass}
m_\phi^2(t) \equiv V^{\prime \prime}\left(\phi_0(t)\right)=k(k-1) \lambda M_P^2\left(\frac{\phi_0(t)}{M_P}\right)^{k-2} \, ,
\end{equation}
and the inflaton energy density can be expressed as
\begin{equation}
    \label{eq:infenden}
    \rho_{\phi}(t) \; = \; V(\phi_0(t)) \; = \; \lambda M_P^4  \left(\frac{\phi_0(t)}{M_P}\right)^{k} \; = \; \frac{1}{k(k-1)} m_{\phi}^2(t) \phi_0^2 \, .
\end{equation}

We use these expressions in following appendix to compute the decay and scattering rates for different inflaton and vector boson couplings, as well as different values of 
$k$.

\section{Amplitudes and Thermal Rates}
\label{ap:ampandthermal}
In this appendix, we calculate the thermal production rate of massive vector fields resulting from the scattering of SM or MSSM states, $\mathrm{SM}/\mathrm{MSSM} + \mathrm{SM}/\mathrm{MSSM} \rightarrow A_\mu+ A_\mu$. These states are assumed to be massless since the initial particle momenta $p_1$ and $p_2$ are typically of order of the inflaton mass $m_{\phi}$ at the beginning of reheating. We first determine the squared amplitudes for different spins of the initial state particles. Following that, the thermal rate is derived using the general expression~(\ref{thrate}). We assume that $4m_A^2 \ll s$ and include a factor of $2$ to account for the production of 2 particles per scattering.

The squared amplitudes are expressed in terms of the Mandelstam variables $s$ and $t$, with
\begin{equation}
    t \; = \; \frac{s}{2} \left(\sqrt{1 - \frac{4m_{A}^2}{s}} \cos{\theta_{13}} - 1 \right) + m_{A}^2 \, ,
\end{equation}
\begin{equation}
    s \; = \; 2E_1 E_2(1- \cos{\theta_{12}}) \, .
\end{equation}
The general squared amplitude for thermal processes involving SM  initial states is given by Eq.~(\ref{Eq:ampscat}), where we include $4$ degrees for 1 complex Higgs doublet, $12$ degrees for $8$ gluons and 4 electroweak bosons, and $45$ degrees for 6 
(anti)quarks with 3 colors, 3 (anti)charged leptons and 3 
neutrinos. For MSSM, we include $98$ degrees for squarks, sleptons, and Higgs bosons, $61$ degrees for quarks, leptons, gauginos and higgsinos, and $12$ degrees for gluons and electroweak bosons. We note that the squared amplitudes include the symmetry factors of both the initial and final states, and this is indicated with an overbar.

When summing over all polarizations, the total squared amplitude of the gravity-mediated  production of massive vector from  massless scalar is given by
\begin{equation}
\begin{aligned}
|\mathcal{\overline{M}}^{0 1}|^2=
\frac{3 \left(m_A^2-t\right)^2 \left(m_A^2 -s-t\right)^2}{4 M_P^4 s^2} \, .
\label{fullM2}
\end{aligned}
\end{equation}
Similarly, the matrix element squared for massive vector production from massless fermions, and from massless gauge bosons are
\begin{equation}
    \begin{aligned}
    \label{amp:vect2}
  &     
|\mathcal{\overline{M}}^{\frac{1}{2} 1}|^2 =-\frac{1}{4M_P^4s^2}\left[12m_A^8-12m_A^6(s+4t)+m_A^4(5s^2+48 st  +72t^2)\right.\\&\left.\qquad \qquad -2 m_A^2(2 s^3+11 s^2 t +30 s t ^2+24t^3)+ t (s+t)(5s^2+12 s t +12 t^2)\right]\, ,
    \end{aligned}
\end{equation}

\begin{equation}
    \begin{aligned}
    \label{amp:vect3}
     &|\mathcal{\overline{M}}^{1 1}|^2 =\frac{\left(m_A^4-2 m_A^2 t+s^2+t (s+t)\right) \left(3 \left(m_A^4-2 m_A^2 t+t (s+t)\right)+s^2\right)}{2 M_P^4 s^2}\,.
    \end{aligned}
\end{equation}
By evaluating the integral \eqref{thrate}, we find that the thermal production rate of massive vector fields can be written as 
\bea
&&
R^T_{1} \; = \; \beta_1 \frac{T^{8}}{M_P^4} 
+ \beta_2 \frac{m_{A}^2 T^6}{M_P^4} + \beta_3 \frac{m_{A}^4 T^4}{M_P^4}  \, .
\eea

We parametrize $\beta_i$, with $i=1,2,3$, as 
\begin{equation}
    \beta_i = N_b x_i + N_f y_i + N_Vz_i \, ,
\end{equation}
where
\begin{equation}
    \begin{aligned}
&        \left\{x_1,y_1,z_1\right\}=\left\{\frac{\pi^3}{108000},\frac{637\pi^3}{20736000},\frac{13\pi^3}{162000}\right\} \, ,
\\&        \left\{x_2,y_2,z_2\right\}=\left\{\frac{\zeta(3)^2} {160\pi^5},\frac{21\zeta(3)^2}{640\pi^5},\frac{7\zeta(3)^2}{60\pi^5}\right\} \, ,
\\&        \left\{x_3,y_3,z_3\right\}=\left\{\frac{1}{15360\pi},\frac{1}{23040\pi},\frac{1}{2880\pi} \right\} \, .
    \end{aligned}
\end{equation}

Then for the SM, with $N_b = 4, N_f = 45$, and $N_V = 12$, we find \begin{equation}\beta_1 \; = \; \frac{5489 \pi^3}{2304000} \, ,\label{Eq:beta1}\end{equation}\begin{equation}    \beta_2 \; = \; \frac{1857\zeta(3)^2}{640 \pi ^5} \, ,\label{Eq:beta2}\end{equation}
\begin{equation}   \beta_3 \; = \; \frac{49}{7680 \pi} \, ,\label{Eq:beta3}\end{equation}and for the MSSM, with $N_b = 98, N_f = 61$, and $N_V = 12$, we find\begin{equation}\beta_1 \; = \; \frac{77641 \pi^3}{20736000} \, ,\label{Eq:mssmbeta1}\end{equation}\begin{equation}    \beta_2 \; = \; \frac{2569\zeta(3)^2}{640 \pi ^5} \, ,\label{Eq:mssmbeta2} \end{equation}\begin{equation}    \beta_3 \; = \; \frac{19}{1440 \pi} \, .\label{Eq:mssmbeta3}\end{equation}


\begin{thebibliography}{99}
  \bibitem{reviews}
   K.~A.~Olive,
  Phys.\ Rept.\  {\bf 190} (1990) 307;
A. D. Linde, {\it Particle  
Physics and
Inflationary Cosmology} (Harwood, Chur, Switzerland, 1990); 
  D.~H.~Lyth and A.~Riotto,
{\it Phys.\ Rep.}  {\bf 314} (1999) 1
[arXiv:hep-ph/9807278];
J.~Martin, C.~Ringeval and V.~Vennin,
  Phys.\ Dark Univ.\  {\bf 5-6}, 75-235 (2014)
  [arXiv:1303.3787 [astro-ph.CO]];
  J.~Martin, C.~Ringeval, R.~Trotta and V.~Vennin,
  JCAP {\bf 1403} (2014) 039
  [arXiv:1312.3529 [astro-ph.CO]];
 J.~Martin,
  Astrophys.\ Space Sci.\ Proc.\  {\bf 45}, 41 (2016)
  [arXiv:1502.05733 [astro-ph.CO]].

  
 

\bibitem{dg}
   A.~D.~Dolgov and A.~D.~Linde,
  Phys.\ Lett.\  {\bf 116B}, 329 (1982);
  L.~F.~Abbott, E.~Farhi and M.~B.~Wise,
  Phys.\ Lett.\  {\bf 117B}, 29 (1982).


\bibitem{nos}
  D.~V.~Nanopoulos, K.~A.~Olive and M.~Srednicki,
  Phys.\ Lett.\ B {\bf 127}, 30 (1983).


\bibitem{Giudice:2000ex}
  G.~F.~Giudice, E.~W.~Kolb and A.~Riotto,
  Phys.\ Rev.\ D {\bf 64} (2001) 023508
  [hep-ph/0005123];
   D.~J.~H.~Chung, E.~W.~Kolb and A.~Riotto,
  Phys.\ Rev.\ D {\bf 60} (1999) 063504
  [hep-ph/9809453].
  
 \bibitem{GMOP}
M.~A.~G.~Garcia, Y.~Mambrini, K.~A.~Olive and M.~Peloso,
Phys. Rev. D \textbf{96}, no.10, 103510 (2017)
[arXiv:1709.01549 [hep-ph]].

  \bibitem{Bernal:2020gzm}
N.~Bernal,
JCAP \textbf{10}, 006 (2020)
[arXiv:2005.08988 [hep-ph]];
A.~Di Marco and G.~Pradisi,
Int. J. Mod. Phys. A \textbf{36}, no.15, 2150095 (2021)
[arXiv:2102.00326 [gr-qc]].


\bibitem{GKMO1}
M.~A.~G.~Garcia, K.~Kaneta, Y.~Mambrini and K.~A.~Olive,
Phys. Rev. D \textbf{101} (2020) no.12, 123507
[arXiv:2004.08404 [hep-ph].




\bibitem{GKMO2}
M.~A.~G.~Garcia, K.~Kaneta, Y.~Mambrini and K.~A.~Olive,
JCAP \textbf{04}, 012 (2021)
[arXiv:2012.10756 [hep-ph]].

\bibitem{Becker:2023tvd}
M.~Becker, E.~Copello, J.~Harz, J.~Lang and Y.~Xu,
[arXiv:2306.17238 [hep-ph]].


\bibitem{Davidson:2000er} 
  S.~Davidson and S.~Sarkar,
  JHEP {\bf 0011}, 012 (2000)
  [hep-ph/0009078].

  \bibitem{Harigaya:2013vwa}
  K.~Harigaya, K.~Mukaida and M.~Yamada,
JHEP \textbf{07} (2019), 059
[arXiv:1901.11027 [hep-ph]];
K.~Harigaya, M.~Kawasaki, K.~Mukaida and M.~Yamada,
Phys. Rev. D \textbf{89} (2014) no.8, 083532
[arXiv:1402.2846 [hep-ph]];
K.~Harigaya and K.~Mukaida,
JHEP \textbf{05}, 006 (2014)
[arXiv:1312.3097 [hep-ph]].

\bibitem{Mukaida:2015ria}
K.~Mukaida and M.~Yamada,
JCAP \textbf{02}, 003 (2016)
[arXiv:1506.07661 [hep-ph]].

\bibitem{Passaglia:2021upk}
S.~Passaglia, W.~Hu, A.~J.~Long and D.~Zegeye,
Phys. Rev. D \textbf{104}, no.8, 083540 (2021)
[arXiv:2108.00962 [hep-ph]].
 
  
\bibitem{GA}
 M.~A.~G.~Garcia and M.~A.~Amin,
  Phys.\ Rev.\ D {\bf 98}, no. 10, 103504 (2018)
  [arXiv:1806.01865 [hep-ph]];

\bibitem{Drees:2021lbm}
M.~Drees and B.~Najjari,
JCAP \textbf{10}, 009 (2021)
[arXiv:2105.01935 [hep-ph]].

\bibitem{Drees:2022vvn}
M.~Drees and B.~Najjari,
[arXiv:2205.07741 [hep-ph]].

\bibitem{Mukaida:2022bbo}
K.~Mukaida and M.~Yamada,
JHEP \textbf{10}, 116 (2022)
[arXiv:2208.11708 [hep-ph]].

\bibitem{no-scale}
E.~Cremmer, S.~Ferrara, C.~Kounnas and D.~V.~Nanopoulos,
  Phys.\ Lett.\ B {\bf 133} (1983) 61;
  A.~B.~Lahanas and D.~V.~Nanopoulos,
  Phys.\ Rept.\  {\bf 145} (1987) 1.

\bibitem{Witten}
E.~Witten,
  Phys.\ Lett.\  {\bf 155B} (1985) 151.

\bibitem{building}
J.~Ellis, M.~A.~G.~Garcia, N.~Nagata, N.~D.~V., K.~A.~Olive and S.~Verner,
Int. J. Mod. Phys. D \textbf{29}, no.16, 2030011 (2020)
[arXiv:2009.01709 [hep-ph]].

\bibitem{eno6}
   J.~Ellis, D.~V.~Nanopoulos and K.~A.~Olive,
  Phys.\ Rev.\ Lett.\  {\bf 111} (2013) 111301 
  [arXiv:1305.1247 [hep-th]].

\bibitem{eno7}
J.~Ellis, D.~V.~Nanopoulos and K.~A.~Olive,
JCAP \textbf{10}, 009 (2013)
[arXiv:1307.3537 [hep-th]].

\bibitem{enov1}
J.~Ellis, D.~V.~Nanopoulos, K.~A.~Olive and S.~Verner,
JHEP \textbf{03}, 099 (2019)
[arXiv:1812.02192 [hep-th]].

  \bibitem{FeKR}
  S.~Ferrara, A.~Kehagias and A.~Riotto,
  Fortsch.\ Phys.\  {\bf 62}, 573 (2014)
  [arXiv:1403.5531 [hep-th]];
S.~Ferrara, A.~Kehagias and A.~Riotto,
  Fortsch.\ Phys.\  {\bf 63}, 2 (2015)
  [arXiv:1405.2353 [hep-th]];
 R.~Kallosh, A.~Linde, B.~Vercnocke and W.~Chemissany,
  JCAP {\bf 1407}, 053 (2014)
  [arXiv:1403.7189 [hep-th]];
 K.~Hamaguchi, T.~Moroi and T.~Terada,
  Phys.\ Lett.\ B {\bf 733}, 305 (2014)
  [arXiv:1403.7521 [hep-ph]];
 J.~Ellis, M.~A.~G.~Garc\'ia, D.~V.~Nanopoulos and K.~A.~Olive,
  JCAP {\bf 1405}, 037 (2014)
  [arXiv:1403.7518 [hep-ph]];
  J.~Ellis, M.~A.~G.~Garcia, D.~V.~Nanopoulos and K.~A.~Olive,
JCAP \textbf{08}, 044 (2014)
[arXiv:1405.0271 [hep-ph]];
J.~Ellis, M.~A.~G.~Garc{\' i}a, D.~V.~Nanopoulos and K.~A.~Olive,
JCAP \textbf{01}, 010 (2015)
[arXiv:1409.8197 [hep-ph]].


\bibitem{egno4}
J.~Ellis, M.~A.~G.~Garcia, D.~V.~Nanopoulos and K.~A.~Olive,
JCAP \textbf{10}, 003 (2015)
[arXiv:1503.08867 [hep-ph]].

\bibitem{eno9}
J.~Ellis, D.~V.~Nanopoulos and K.~A.~Olive,
Phys. Rev. D \textbf{97}, no.4, 043530 (2018)
[arXiv:1711.11051 [hep-th]].

\bibitem{enov4}
J.~Ellis, D.~V.~Nanopoulos, K.~A.~Olive and S.~Verner,
JCAP \textbf{08}, 037 (2020)
[arXiv:2004.00643 [hep-ph]].

\bibitem{Staro}
A.~A.~Starobinsky,
  Phys.\ Lett.\ B {\bf 91}, 99 (1980).





\bibitem{ekoty}
M.~Endo, K.~Kadota, K.~A.~Olive, F.~Takahashi and T.~T.~Yanagida,
JCAP \textbf{02}, 018 (2007)
[arXiv:hep-ph/0612263 [hep-ph]].

  \bibitem{eno8}
  J.~Ellis, D.~V.~Nanopoulos and K.~A.~Olive,
  Phys.\ Rev.\ D {\bf 89} (2014) 4,  043502
  [arXiv:1310.4770 [hep-ph]];

 \bibitem{snu}
  H.~Murayama, H.~Suzuki, T.~Yanagida and J.-i.~Yokoyama,
  Phys.\ Rev.\ Lett.\  {\bf 70} (1993) 1912 and
  Phys.\ Rev.\ D {\bf 50} (1994) 2356
  [hep-ph/9311326];
  J.~R.~Ellis, M.~Raidal and T.~Yanagida,
  Phys.\ Lett.\ B {\bf 581} (2004) 9
  [hep-ph/0303242];
  D.~Croon, J.~Ellis and N.~E.~Mavromatos,
  Phys.\ Lett.\ B {\bf 724} (2013) 165
  [arXiv:1303.6253 [astro-ph.CO]];
   K.~Nakayama, F.~Takahashi and T.~T.~Yanagida,
  Phys.\ Lett.\ B {\bf 730}, 24 (2014)
  [arXiv:1311.4253 [hep-ph]];
  J.~Ellis, N.~E.~Mavromatos and D.~J.~Mulryne,
  JCAP {\bf 1405} (2014) 012
  [arXiv:1401.6078 [astro-ph.CO]];
J.~L.~Evans, T.~Gherghetta and M.~Peloso,
Phys. Rev. D \textbf{92}, no.2, 021303 (2015)
[arXiv:1501.06560 [hep-ph]].



 \bibitem{klor}
  R.~Kallosh, A.~Linde, K.~A.~Olive and T.~Rube,
  Phys.\ Rev.\ D {\bf 84}, 083519 (2011)
  [arXiv:1106.6025 [hep-th]].



\bibitem{Terada:2014uia}
T.~Terada, Y.~Watanabe, Y.~Yamada and J.~Yokoyama,
JHEP \textbf{02}, 105 (2015)
[arXiv:1411.6746 [hep-ph]].

  \bibitem{dlmmo}
   E.~Dudas, A.~Linde, Y.~Mambrini, A.~Mustafayev and K.~A.~Olive,
  Eur.\ Phys.\ J.\ C {\bf 73} (2013) 2268
  [arXiv:1209.0499 [hep-ph]].

\bibitem{Bernal:2018qlk}
  N.~Bernal, M.~Dutra, Y.~Mambrini, K.~Olive, M.~Peloso and M.~Pierre,
  Phys.\ Rev.\ D {\bf 97} (2018) no.11,  115020
  [arXiv:1803.01866 [hep-ph]].



  
\bibitem{MO}
Y.~Mambrini and K.~A.~Olive,
Phys. Rev. D \textbf{103} (2021) no.11, 115009
[arXiv:2102.06214 [hep-ph]]. 

\bibitem{Bernal:2021kaj}
N.~Bernal and C.~S.~Fong,
JCAP \textbf{06} (2021), 028
[arXiv:2103.06896 [hep-ph]];
X.~Sun,
[arXiv:2112.04304 [hep-ph]].

\bibitem{Barman:2021ugy}
B.~Barman and N.~Bernal,
JCAP \textbf{06}, 011 (2021)
[arXiv:2104.10699 [hep-ph]].



\bibitem{cmov}
S.~Clery, Y.~Mambrini, K.~A.~Olive and S.~Verner,
Phys. Rev. D \textbf{105}, no.7, 075005 (2022)
[arXiv:2112.15214 [hep-ph]].

\bibitem{Haque:2022kez}
M.~R.~Haque and D.~Maity,
Phys. Rev. D \textbf{107}, no.4, 043531 (2023)
[arXiv:2201.02348 [hep-ph]].

\bibitem{cmosv}
S.~Clery, Y.~Mambrini, K.~A.~Olive, A.~Shkerin and S.~Verner,
Phys. Rev. D \textbf{105}, no.9, 095042 (2022)
[arXiv:2203.02004 [hep-ph]].

\bibitem{cmo}
R.~T.~Co, Y.~Mambrini and K.~A.~Olive,
Phys. Rev. D \textbf{106}, no.7, 075006 (2022)
[arXiv:2205.01689 [hep-ph]].

\bibitem{Barman:2022qgt}
B.~Barman, S.~Cl\'ery, R.~T.~Co, Y.~Mambrini and K.~A.~Olive,
JHEP \textbf{12}, 072 (2022)
[arXiv:2210.05716 [hep-ph]].

\bibitem{kkmov}
K.~Kaneta, W.~Ke, Y.~Mambrini, K.~A.~Olive and S.~Verner,
[arXiv:2309.15146 [hep-ph]].




\bibitem{ema}
Y.~Ema, R.~Jinno, K.~Mukaida and K.~Nakayama,
JCAP \textbf{05}, 038 (2015)
[arXiv:1502.02475 [hep-ph]];
Y.~Ema, R.~Jinno, K.~Mukaida and K.~Nakayama,
Phys. Rev. D \textbf{94}, no.6, 063517 (2016)
[arXiv:1604.08898 [hep-ph]];
Y.~Ema, K.~Nakayama and Y.~Tang,
JHEP \textbf{09}, 135 (2018)
[arXiv:1804.07471 [hep-ph]].

\bibitem{Dimopoulos:2006ms}
K.~Dimopoulos,
Phys. Rev. D \textbf{74}, 083502 (2006)
[arXiv:hep-ph/0607229 [hep-ph]].

\bibitem{Graham:2015rva}
P.~W.~Graham, J.~Mardon and S.~Rajendran,
Phys. Rev. D \textbf{93}, no.10, 103520 (2016)
[arXiv:1504.02102 [hep-ph]].

\bibitem{Garny:2015sjg}
M.~Garny, M.~Sandora and M.~S.~Sloth,
Phys. Rev. Lett. \textbf{116}, no.10, 101302 (2016)
[arXiv:1511.03278 [hep-ph]];
M.~Garny, A.~Palessandro, M.~Sandora and M.~S.~Sloth,
JCAP \textbf{02}, 027 (2018)
[arXiv:1709.09688 [hep-ph]].


\bibitem{Tang:2017hvq}
Y.~Tang and Y.~L.~Wu,
Phys. Lett. B \textbf{774}, 676-681 (2017)
[arXiv:1708.05138 [hep-ph]].

\bibitem{Ema:2019yrd}
Y.~Ema, K.~Nakayama and Y.~Tang,
JHEP \textbf{07}, 060 (2019)
[arXiv:1903.10973 [hep-ph]].


\bibitem{Chianese:2020yjo}
M.~Chianese, B.~Fu and S.~F.~King,
JCAP \textbf{06}, 019 (2020)
[arXiv:2003.07366 [hep-ph]];
M.~Chianese, B.~Fu and S.~F.~King,
JCAP \textbf{01}, 034 (2021)
[arXiv:2009.01847 [hep-ph]].

\bibitem{Ahmed:2020fhc}
A.~Ahmed, B.~Grzadkowski and A.~Socha,
JHEP \textbf{08}, 059 (2020)
[arXiv:2005.01766 [hep-ph]].

\bibitem{Kolb:2020fwh}
E.~W.~Kolb and A.~J.~Long,
JHEP \textbf{03}, 283 (2021)
[arXiv:2009.03828 [astro-ph.CO]].

\bibitem{Redi:2020ffc}
M.~Redi, A.~Tesi and H.~Tillim,
JHEP \textbf{05}, 010 (2021)
[arXiv:2011.10565 [hep-ph]].

\bibitem{Ling:2021zlj}
S.~Ling and A.~J.~Long,
Phys. Rev. D \textbf{103}, no.10, 103532 (2021)
[arXiv:2101.11621 [astro-ph.CO]].

\bibitem{Ahmed:2021fvt}
A.~Ahmed, B.~Grzadkowski and A.~Socha,
Phys. Lett. B \textbf{831}, 137201 (2022)
[arXiv:2111.06065 [hep-ph]].

\bibitem{Haque:2021mab}
M.~R.~Haque and D.~Maity,
Phys. Rev. D \textbf{106}, no.2, 023506 (2022)
doi:10.1103/PhysRevD.106.023506
[arXiv:2112.14668 [hep-ph]].







\bibitem{Aoki:2022dzd}
S.~Aoki, H.~M.~Lee, A.~G.~Menkara and K.~Yamashita,
JHEP \textbf{05}, 121 (2022)
[arXiv:2202.13063 [hep-ph]].



\bibitem{Ahmed:2022tfm}
A.~Ahmed, B.~Grzadkowski and A.~Socha,
JHEP \textbf{02}, 196 (2023)
[arXiv:2207.11218 [hep-ph]].

\bibitem{Basso:2022tpd}
E.~Basso, D.~J.~H.~Chung, E.~W.~Kolb and A.~J.~Long,
JHEP \textbf{12}, 108 (2022)
[arXiv:2209.01713 [gr-qc]].

\bibitem{Haque:2023yra}
M.~R.~Haque, D.~Maity and R.~Mondal,
JHEP \textbf{09}, 012 (2023)
[arXiv:2301.01641 [hep-ph]].

\bibitem{Kaneta:2022gug}
K.~Kaneta, S.~M.~Lee and K.~y.~Oda,
JCAP \textbf{09}, 018 (2022)
[arXiv:2206.10929 [astro-ph.CO]].

\bibitem{Kolb:2023dzp}
E.~W.~Kolb, S.~Ling, A.~J.~Long and R.~A.~Rosen,
JHEP \textbf{05}, 181 (2023)
[arXiv:2302.04390 [astro-ph.CO]].

\bibitem{Kaneta:2023kfv}
K.~Kaneta and K.~y.~Oda,
[arXiv:2304.12578 [hep-ph]].

\bibitem{Garcia:2023qab}
M.~A.~G.~Garcia, M.~Pierre and S.~Verner,
[arXiv:2305.14446 [hep-ph]].

\bibitem{Zhang:2023xcd}
R.~Zhang, Z.~Xu and S.~Zheng,
JCAP \textbf{07}, 048 (2023)
[arXiv:2305.02568 [hep-ph]].

\bibitem{RiajulHaque:2023cqe}
M.~Riajul Haque, E.~Kpatcha, D.~Maity and Y.~Mambrini,
Phys. Rev. D \textbf{108} (2023) no.6, 063523
[arXiv:2305.10518 [hep-ph]].

\bibitem{Ozsoy:2023gnl}
O.~\"Ozsoy and G.~Tasinato,
[arXiv:2310.03862 [astro-ph.CO]].

\bibitem{Cembranos:2023qph}
J.~A.~R.~Cembranos, L.~J.~Garay, \'A.~Parra-L\'opez and J.~M.~S\'anchez Vel\'azquez,
[arXiv:2310.07515 [gr-qc]].

\bibitem{Figueroa:2018twl}
D.~G.~Figueroa and E.~H.~Tanin,
JCAP \textbf{10}, 050 (2019)
[arXiv:1811.04093 [astro-ph.CO]].

\bibitem{egnov}
J.~Ellis, M.~A.~G.~Garcia, D.~V.~Nanopoulos, K.~A.~Olive and S.~Verner,
Phys. Rev. D \textbf{105} (2022) no.4, 043504
[arXiv:2112.04466 [hep-ph]].

 \bibitem{Planck}
  N.~Aghanim \textit{et al.} [Planck],
Astron. Astrophys. \textbf{641}, A6 (2020)
[arXiv:1807.06209 [astro-ph.CO]].

\bibitem{Barnaby-Namba-Peloso}
N.~Barnaby, R.~Namba and M.~Peloso,
JCAP \textbf{04}, 009 (2011)
[arXiv:1102.4333 [astro-ph.CO]];
Phys. Rev. D \textbf{85}, 123523 (2012)
[arXiv:1202.1469 [astro-ph.CO]].

\bibitem{Shtanov:1994ce}
Y.~Shtanov, J.~H.~Traschen and R.~H.~Brandenberger,
Phys. Rev. D \textbf{51} (1995), 5438-5455
[arXiv:hep-ph/9407247 [hep-ph]].



\bibitem{Ichikawa:2008ne}
K.~Ichikawa, T.~Suyama, T.~Takahashi and M.~Yamaguchi,
Phys. Rev. D \textbf{78} (2008), 063545
[arXiv:0807.3988 [astro-ph]].

\bibitem{Kallosh:2013hoa}
R.~Kallosh and A.~Linde,
JCAP \textbf{07}, 002 (2013)
[arXiv:1306.5220 [hep-th]].

\bibitem{Amin:2011hj}
M.~A.~Amin, R.~Easther, H.~Finkel, R.~Flauger and M.~P.~Hertzberg,
Phys. Rev. Lett. \textbf{108}, 241302 (2012)
[arXiv:1106.3335 [astro-ph.CO]].


\bibitem{Lozanov:2016hid}
K.~D.~Lozanov and M.~A.~Amin,
Phys. Rev. Lett. \textbf{119}, no.6, 061301 (2017)
[arXiv:1608.01213 [astro-ph.CO]].

\bibitem{Lozanov:2017hjm}
K.~D.~Lozanov and M.~A.~Amin,
Phys. Rev. D \textbf{97}, no.2, 023533 (2018)
[arXiv:1710.06851 [astro-ph.CO]].

\bibitem{Antusch:2021aiw}
S.~Antusch, D.~G.~Figueroa, K.~Marschall and F.~Torrenti,
Phys. Rev. D \textbf{105}, no.4, 043532 (2022)
[arXiv:2112.11280 [astro-ph.CO]].

\bibitem{Garcia:2023eol}
M.~A.~G.~Garcia and M.~Pierre,
JCAP \textbf{11}, 004 (2023)
[arXiv:2306.08038 [hep-ph]].

\bibitem{Garcia:2023dyf}
M.~A.~G.~Garcia, M.~Gross, Y.~Mambrini, K.~A.~Olive, M.~Pierre and J.~H.~Yoon,
[arXiv:2308.16231 [hep-ph]].


\bibitem{hol}
B.~R.~Holstein,
Am. J. Phys. \textbf{74}, 1002-1011 (2006)
[arXiv:gr-qc/0607045 [gr-qc]].

\bibitem{mybook} Y. Mambrini, Particles in the dark Universe, {\it Springer Ed., ISBN 978-3-030-78139-2 (2021)}.


\bibitem{Barman:2022tzk}
B.~Barman, N.~Bernal, Y.~Xu and \'O.~Zapata,
JCAP \textbf{07} (2022) no.07, 019
[arXiv:2202.12906 [hep-ph]].

\bibitem{Kainulainen:2016vzv}
K.~Kainulainen, S.~Nurmi, T.~Tenkanen, K.~Tuominen and V.~Vaskonen,
JCAP \textbf{06} (2016), 022
[arXiv:1601.07733 [astro-ph.CO]].


\bibitem{Benakli:2017whb}
K.~Benakli, Y.~Chen, E.~Dudas and Y.~Mambrini,
Phys. Rev. D \textbf{95}, no.9, 095002 (2017)
[arXiv:1701.06574 [hep-ph]].

  \bibitem{grav2}
  E.~Dudas, Y.~Mambrini and K.~Olive,
  Phys.\ Rev.\ Lett.\  {\bf 119} (2017) no.5,  051801
  [arXiv:1704.03008 [hep-ph]].

















\bibitem{Stueckelberg:1938hvi}
E.~C.~G.~Stueckelberg,
Helv. Phys. Acta \textbf{11}, 225-244 (1938)
doi:10.5169/seals-110852

\bibitem{Ruegg:2003ps}
H.~Ruegg and M.~Ruiz-Altaba,
Int. J. Mod. Phys. A \textbf{19}, 3265-3348 (2004)
[arXiv:hep-th/0304245 [hep-th]].

\bibitem{Kors:2004dx}
B.~Kors and P.~Nath,
Phys. Lett. B \textbf{586}, 366-372 (2004)
[arXiv:hep-ph/0402047 [hep-ph]].
\bibitem{Kors:2004iz}
B.~Kors and P.~Nath,
doi:10.1142/9789812701756\_0056
[arXiv:hep-ph/0411406 [hep-ph]].


\bibitem{Kors:2005uz}
B.~Kors and P.~Nath,
JHEP \textbf{07}, 069 (2005)
[arXiv:hep-ph/0503208 [hep-ph]].


\bibitem{Belokogne:2015etf}
A.~Belokogne and A.~Folacci,
Phys. Rev. D \textbf{93}, no.4, 044063 (2016)
[arXiv:1512.06326 [gr-qc]].

\bibitem{Nurmi:2015ema}
S.~Nurmi, T.~Tenkanen and K.~Tuominen,
JCAP \textbf{11}, 001 (2015)
doi:10.1088/1475-7516/2015/11/001
[arXiv:1506.04048 [astro-ph.CO]].



\end{thebibliography}
\end{document}